\documentclass[preprintnumbers,a4paper,11pt]{article}
\pdfoutput=1
\usepackage[utf8]{inputenc}
\usepackage{jheppub} 

\usepackage{footmisc,enumerate}
\usepackage{multirow}
\usepackage{subfigure}
\usepackage{amsmath,amssymb}
\usepackage{graphicx}
\usepackage{placeins}
\usepackage{bbm}
\usepackage{xspace,slashed}
\usepackage{hyperref}
\hypersetup{colorlinks=true, citecolor=blue, urlcolor=blue, linkcolor=blue}
\usepackage[normalem]{ulem}
\usepackage{amsmath}
\usepackage{amsfonts}
\usepackage{amssymb}
\usepackage{graphicx}
\usepackage{cancel}
\usepackage{xcolor}
\usepackage{hyperref}
\usepackage{multirow}
\usepackage{pifont}

\usepackage[autostyle]{csquotes}

\renewcommand{\vec}[1]{{\bf{#1}}}

\begin{document}

\newcommand{\og}{\ensuremath{\tilde{O}_g}\xspace}
\newcommand{\ot}{\ensuremath{\tilde{O}_t}\xspace}

\providecommand{\abs}[1]{\lvert#1\rvert}

\newcommand{\cw}{\ensuremath{C_{\widetilde{W}}\xspace}}
\newcommand{\chwb}{\ensuremath{C_{H\widetilde{W}B}}\xspace}

\title{Anomaly detection with Convolutional Graph Neural Networks}

\preprint{IPPP/20/102}

\author[a]{Oliver Atkinson,}

\author[a]{Akanksha~Bhardwaj,} 
\author[a]{Christoph~Englert,} 
\author[b,c]{Vishal~S.~Ngairangbam,}

\author[d,e]{and Michael~Spannowsky} 

\affiliation[a]{School of Physics \& Astronomy, University of Glasgow, Glasgow G12 8QQ, United Kingdom}
\affiliation[b]{Theoretical Physics Division, Physical Research Laboratory, Shree Pannalal Patel Marg, Ahmedabad - 380009, Gujarat, India}
\affiliation[c]{Discipline of Physics, Indian Institute of Technology, Palaj, Gandhinagar - 382424, Gujarat, India}
\affiliation[d]{Institute for Particle Physics Phenomenology, Durham University, Durham DH1 3LE, United Kingdom}
\affiliation[e]{Department of Physics, Durham University, Durham DH1 3LE, United Kingdom}

\emailAdd{o.atkinson.1@research.gla.ac.uk} 
\emailAdd{akanksha.bhardwaj@glasgow.ac.uk}
\emailAdd{christoph.englert@glasgow.ac.uk}
\emailAdd{vishalng@prl.res.in}
\emailAdd{michael.spannowsky@durham.ac.uk}
\preprint{}
\abstract{
	We devise an autoencoder based strategy to facilitate anomaly detection for boosted jets, employing Graph Neural Networks (GNNs) to do so. To overcome known limitations of GNN autoencoders, we design a symmetric decoder capable of simultaneously reconstructing edge features and node features. 
	Focusing on latent space based discriminators, we find that such setups provide a promising avenue to isolate new physics and competing SM signatures from sensitivity-limiting QCD jet contributions. We demonstrate the flexibility and broad applicability of this approach using examples of $W$ bosons, top quarks, and exotic hadronically-decaying exotic scalar bosons.
}
\maketitle

\section{Introduction}
\label{sec:intro}
The search for new physics beyond the SM (BSM), albeit unsuccessful so far at high energy colliders such as the Large Hadron Collider (LHC), remains a key pillar of the particle physics phenomenology programme. The mounting pressure of exclusion constraints for well-motivated BSM scenarios reported by the ATLAS and CMS experiments has shifted theoretical efforts more towards model-independent approaches for new physics discoveries. This is most prominently reflected in the recent resurgence of effective field theory interpretations of collider data~(see e.g. Ref.~\cite{Brivio:2017vri} for a recent review). 

One method of searching for the presence of new BSM interactions is attempting to detect anomalies in otherwise well-understood and abundant collider data~\cite{Collins:2021nxn,CMS:2020zjg,Aaboud:2018ufy,Collins:2018epr,Blance:2019ibf,Hajer:2018kqm,DeSimone:2018efk,Nachman:2020lpy,Nachman:2020ccu}. Hard jets produced primarily via QCD-mediated interactions at hadron colliders are a well-known laboratory for such strategies and their phenomenology has seen considerable development over the past decade~\cite{Marzani:2019hun}. QCD jets are produced with large rates at the LHC even when they are extremely hard, and therefore are the typical backgrounds that any search for new physics in hadronic final states needs to overcome. Turning this argument around, we can classify jets, possibly using motivated first-principle approaches~\cite{Soper:2011cr,Soper:2014rya}, and isolate more interesting generic BSM-type signatures by vetoing a ``typical'' QCD jet. Following this line of thought, anomaly-detection has become a primary application of unsupervised machine learning. This typically involves so-called autoencoders, which are artificial neural networks specifically tailored to reproduce the most common properties of a training data set via a reduction of the dimensionality of the input's features. When a jet behaves less like a common QCD jet (e.g. in case of a particular hadronic BSM decay) such a network should perform poorly, i.e. the loss that parametrises the networks ability to reproduce the QCD signature can be used as a BSM-discriminating observable.

The evolution of a typical QCD event from high to low energies is well understood over a vast range of energy scales, as demonstrated by the successful application of QCD shower Monte Carlo programmes to the modelling of collider data~(see e.g.~\cite{Aad:2014qxa}). This evolution also motivates the application of Graph Neural Networks (GNNs)~\cite{zhou2018graph,9046288} to QCD phenomenology as recently done in~Refs.~\cite{Dreyer:2020brq}, and also to  exploiting the Lund-plane representation of splittings~\cite{Andersson:1988gp,Lifson:2020gua}. GNNs have also been studied in various scenarios~\cite{Mikuni:2020wpr,Knapp:2020dde,Mikuni:2020qds,Dezoort:2021kfk,Abdughani:2018wrw} at the LHC. Moreover, they have also shown promising performances for use in real-time triggers~\cite{Iiyama:2020wap}.
In this work we consider a GNN-based autoencoder for anomaly detection in boosted QCD jets data. Convolutional autoencoders have been proposed and studied in~\cite{Farina:2018fyg,Heimel:2018mkt,Roy:2019jae,Khosa:2020qrz,Finke:2021sdf,Cheng:2020dal} for distinguishing QCD jets from non-QCD jets using ``jet-images"~\cite{Cogan:2014oua,deOliveira:2015xxd} as the input space. However, convolutions on these images are expensive due to their extreme sparsity. Moreover, CNNs, in principle, are limited to the Euclidean domain. GNNs mitigate these two inadequacies, so studying their performance as anomaly finders is motivated. Supervised jet classification with GNNs has been studied in Ref.~\cite{Ju:2020tbo,Qu:2019gqs,Dreyer:2020brq}, while unsupervised clustering of event-graphs with photonic quantum computers have been explored in Ref.~\cite{Blance:2021gcs}.A study of particle graph autoencoders for anomaly detection has been carried out in the LHC Olympics community challenge~\cite{kasieczka2021lhc}. A typical obstacle of GNN-based autoencoders is achieving an appropriate reflection of all network features on the decoding side. Existing graph-autoencoders in the literature~\cite{kipf2016variational,tran2018learning,salha2020simple,pan2018adversarially,park2019symmetric} are designed mostly for node-classification or link prediction, while we desire a network capable of classifying graphs. Moreover, jets provide us with multidimensional edge information, along with node features; classifying the entire graph thereby exploits the full kinematic information of the event. To solve this known difficulty of graph-autoencoders, we design a decoder network capable of simultaneously reconstructing multidimensional edge, and node features with the help of \emph{Inner Product Layers}.

This paper is organised as follows: Sec.~\ref{sec:details} introduces our analysis setup. The graph neural network methodology that we use in this work is described in Sec.~\ref{sec:net_arch}, where we provide details on the network's architecture and its performance. Results are presented in Sect.~\ref{sec:results}, and we conclude in Sec.~\ref{sec:conc}.

\section{Elements of the Simulation}
\label{sec:details}
For our proof-of-principle analysis\footnote{Throughout this work, we will focus on 13 TeV LHC collisions.}, we generate events using {\tt MadGraph5}~\cite{Alwall:2014hca} at leading order (LO), followed by {\tt Pythia8}~\cite{Sjostrand:2006za} for showering and hadronization. The hadronic final states are then clustered into jets using the anti-$k_t$ algorithm~\cite{Cacciari:2008gp} with parameter $R=1.5$ using {\tt FastJet}~\cite{Cacciari:2011ma}. Along with a requirement that the rapidity of jets is $|y| < 2.5$, the minimum transverse momentum of a jet is required to be $p_T > 1~\text{TeV}$ for this ``fat jet'' cluster. Only the leading jet from each multi-jet event is used as an input to the graph network and we do not include detector effects to our analysis. The sample used for training of the autoencoder (for details see below) is a QCD multi-jet background sample, consisting of 200k generated $pp \to jj$ events.   

To test the autoencoder's anomaly detection performance we use three different signal samples, each consisting of 100k events generated with {\tt MadGraph5}, using the same procedure described above. These samples consist of 
\begin{enumerate}[(i)]
	\item boosted hadronically-decaying $W$ bosons as a benchmark for two-prong jet structure, 
	\item boosted hadronically-decaying top quarks, as a benchmark for a three-prong structure, and 
	\item a boosted scalar $\phi$ decaying as $\phi \to W^+W^- \to 4 j$ to give a four-prong structure. The interaction is based on a simplified Lagrangian
	\begin{equation}
	\mathcal{L} \supset -\frac{c_1}{v} \phi W^{\mu \nu} W_{\mu \nu} - c_2  (u \bar{u} + d\bar{d})\phi,
	\end{equation}
	where $c_1$ and $c_2$ are dimensionless constants and $v$ is the Higgs field's vacuum expectation value (vev). We choose $m_\phi=700~\text{GeV}$ for demonstration purposes, but note that our results are not too sensitive to the $\phi$ mass scale.
\end{enumerate}

\begin{figure*}[t]
	\centering
	\includegraphics[scale=0.24]{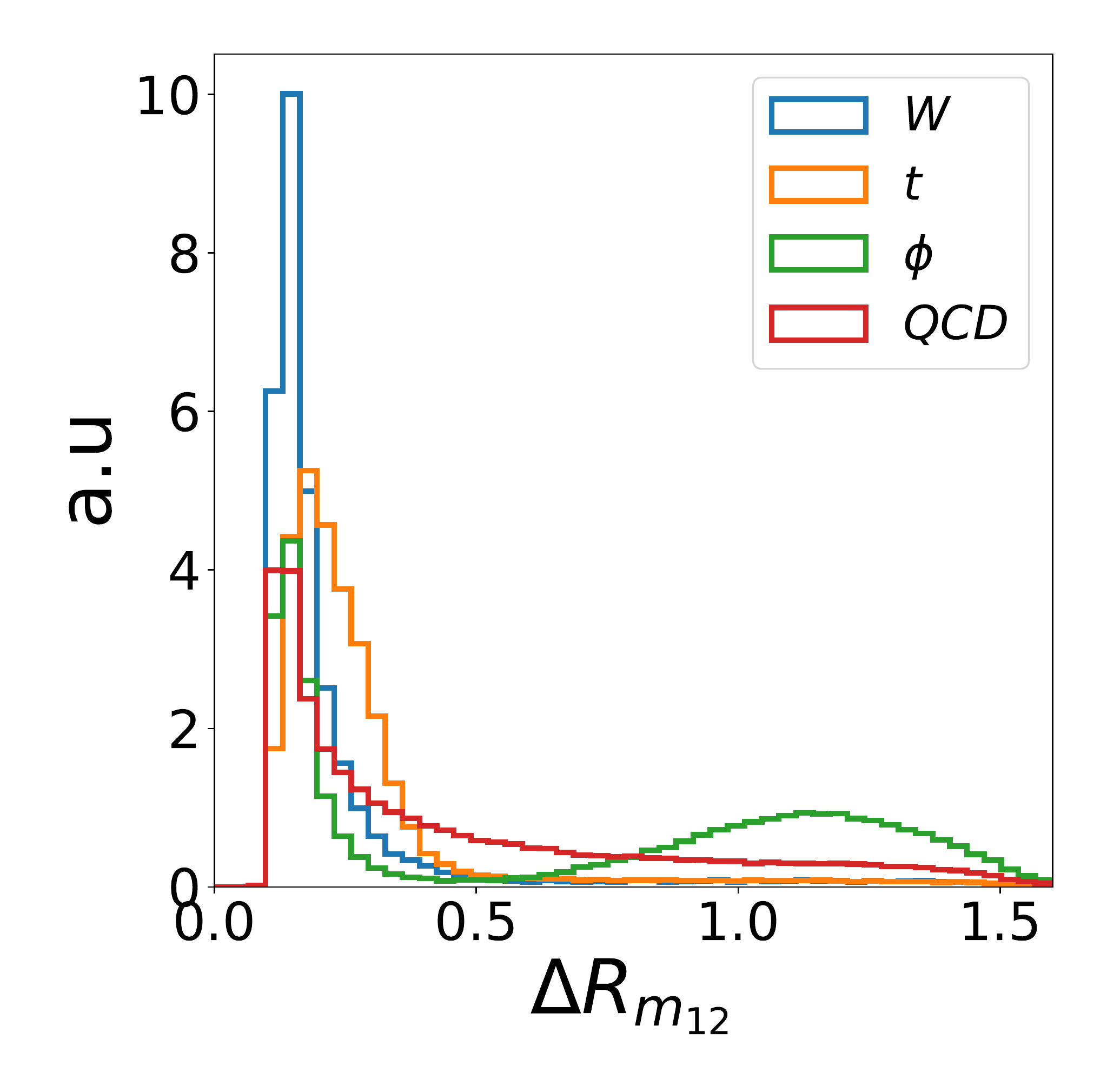}
	\includegraphics[scale=0.24]{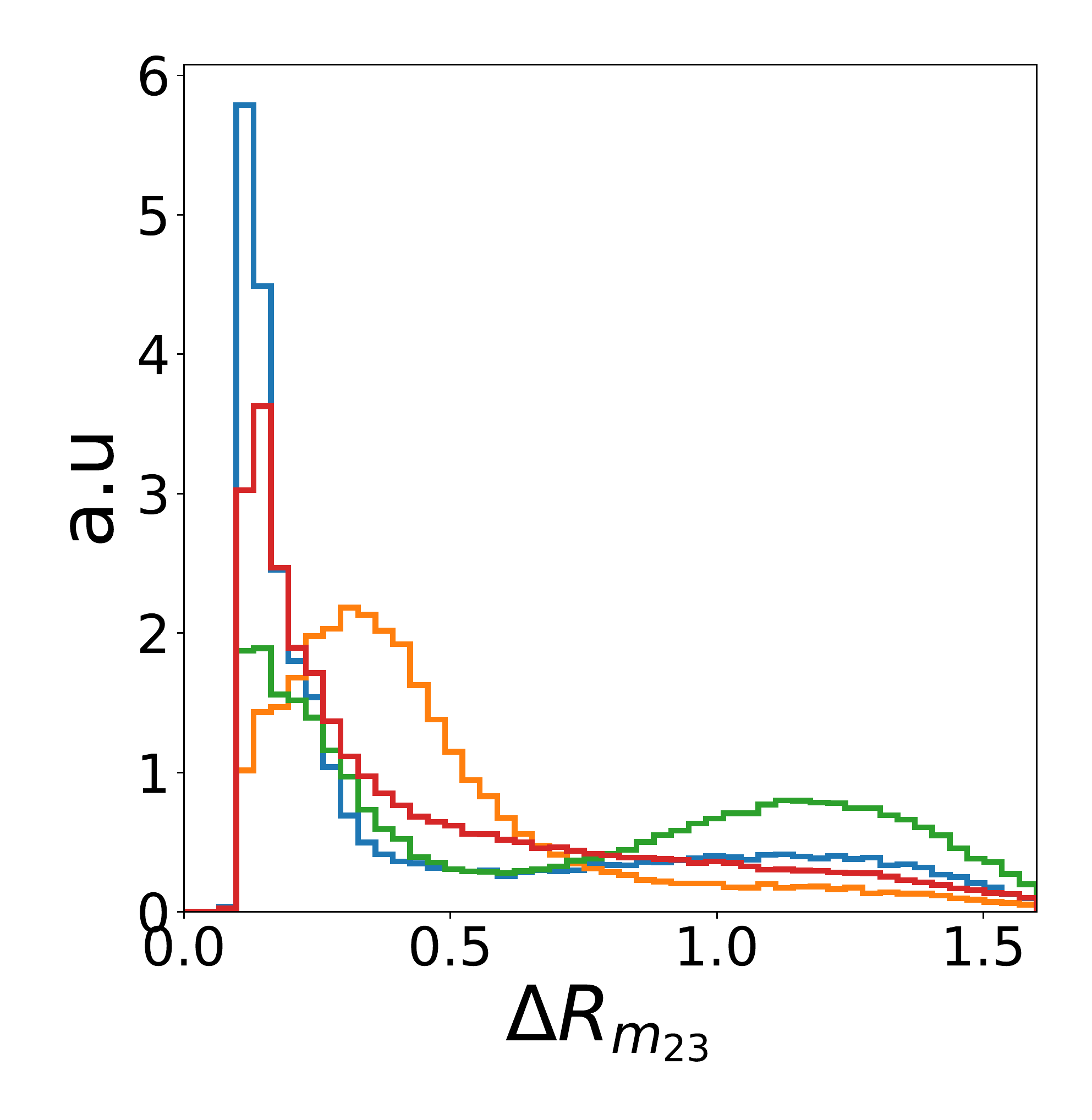}
	\includegraphics[scale=0.24]{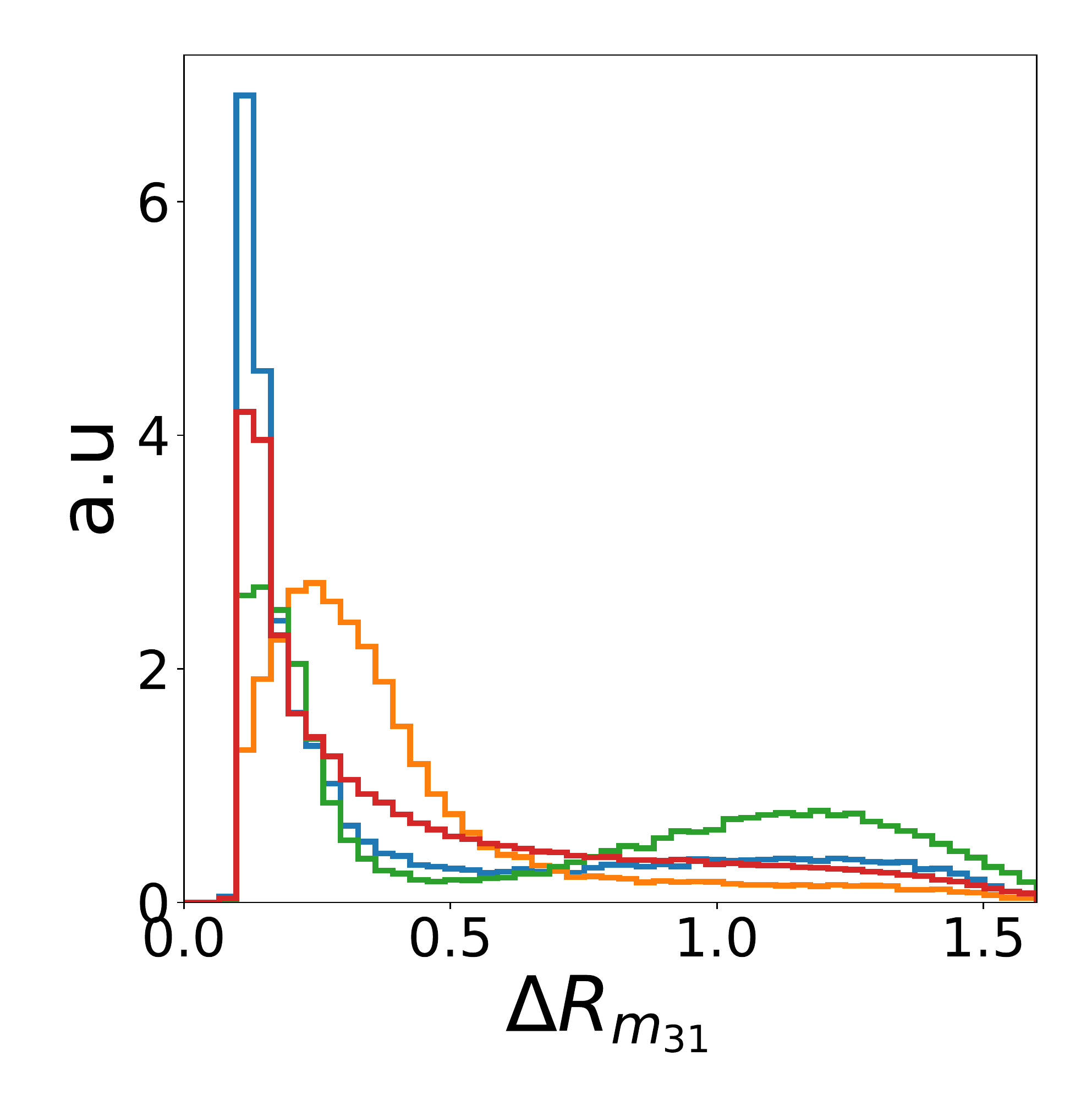}
	\caption{Normalised angular separation distribution between three leading microjets in the fat jet for the physics scenarios discussed in this work.}
	\label{fig:microjetsR}
\end{figure*}

\begin{figure*}[t]
	\centering
	\includegraphics[scale=0.21]{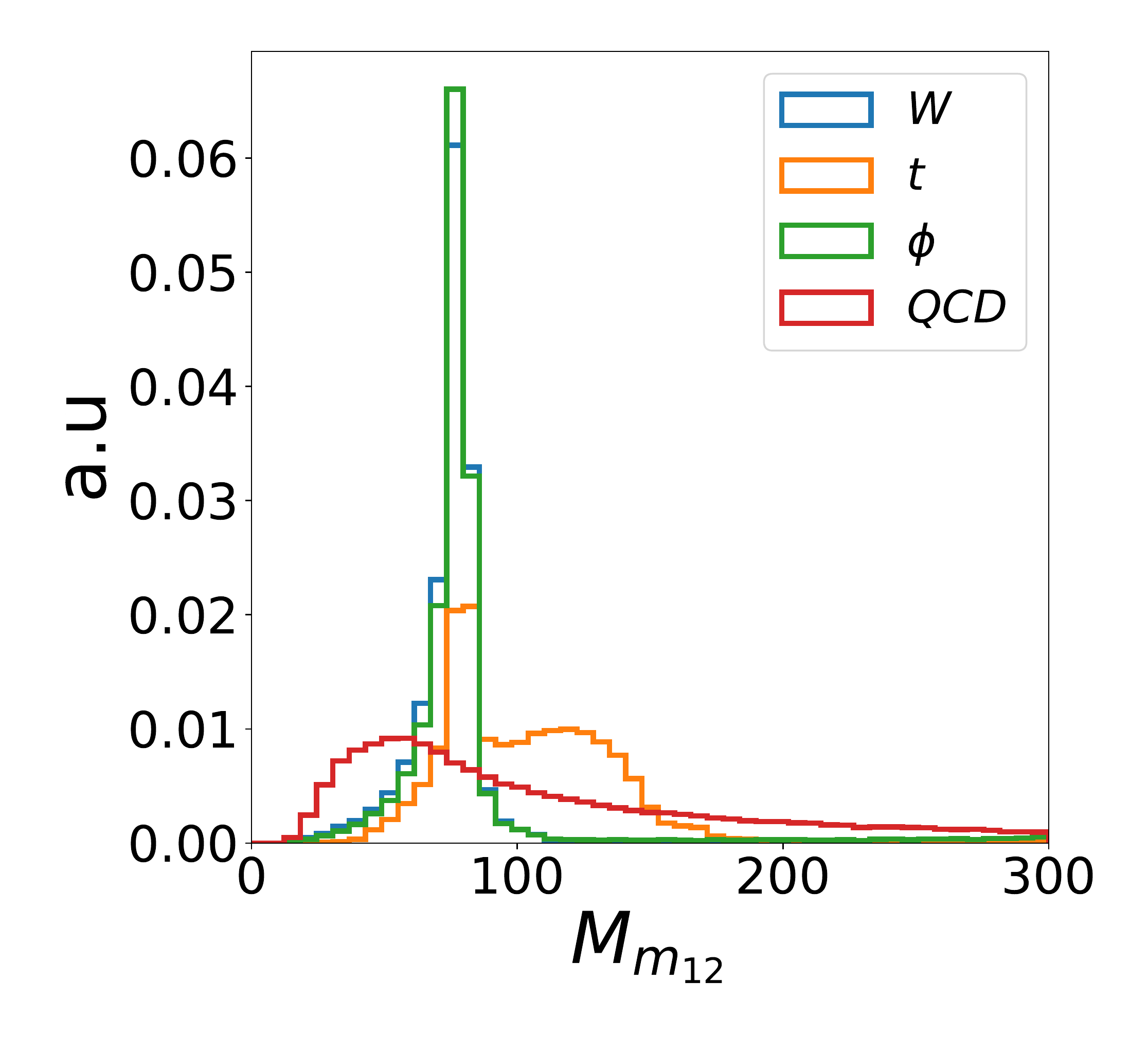}
	\includegraphics[scale=0.21]{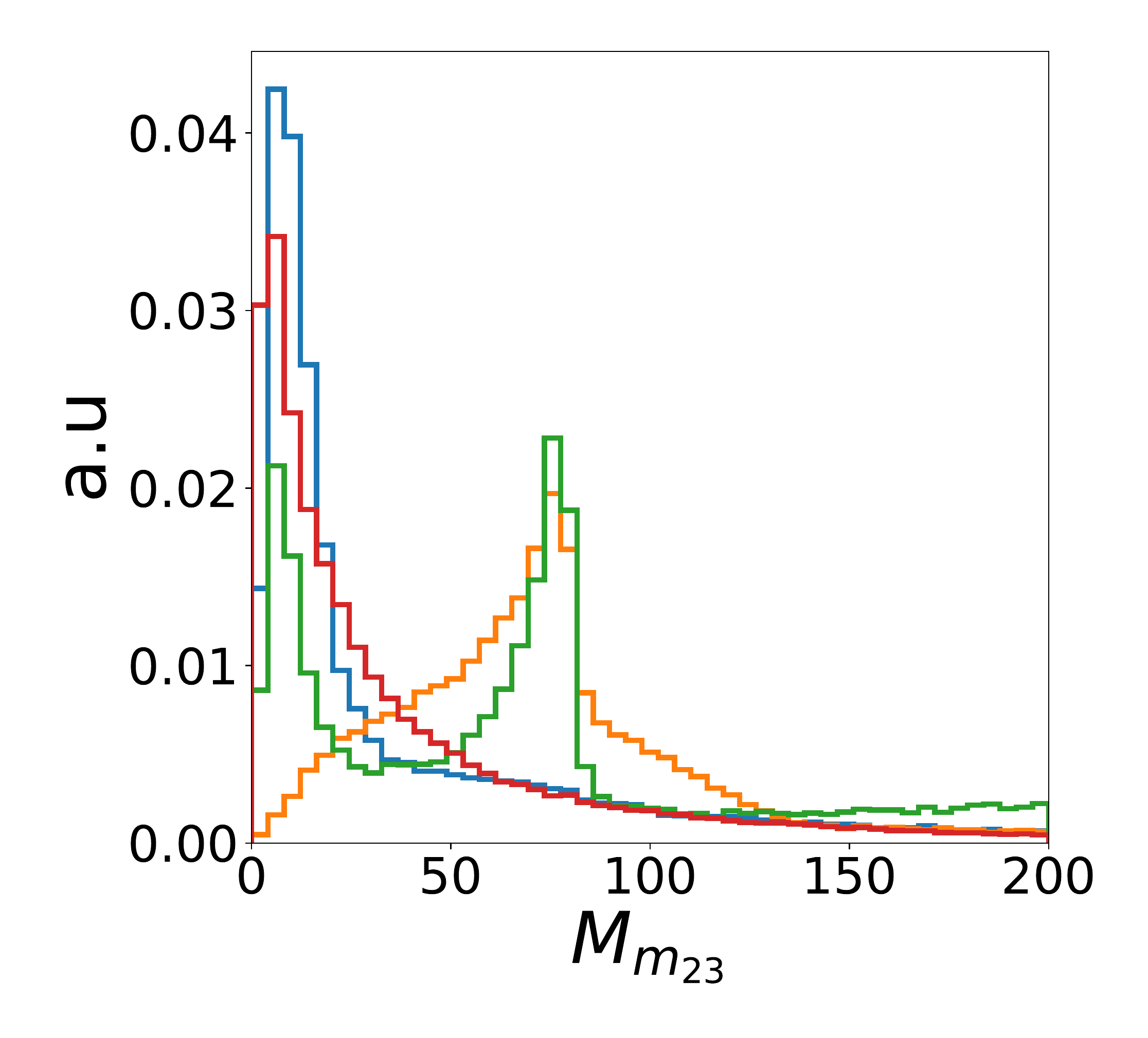}
	\includegraphics[scale=0.21]{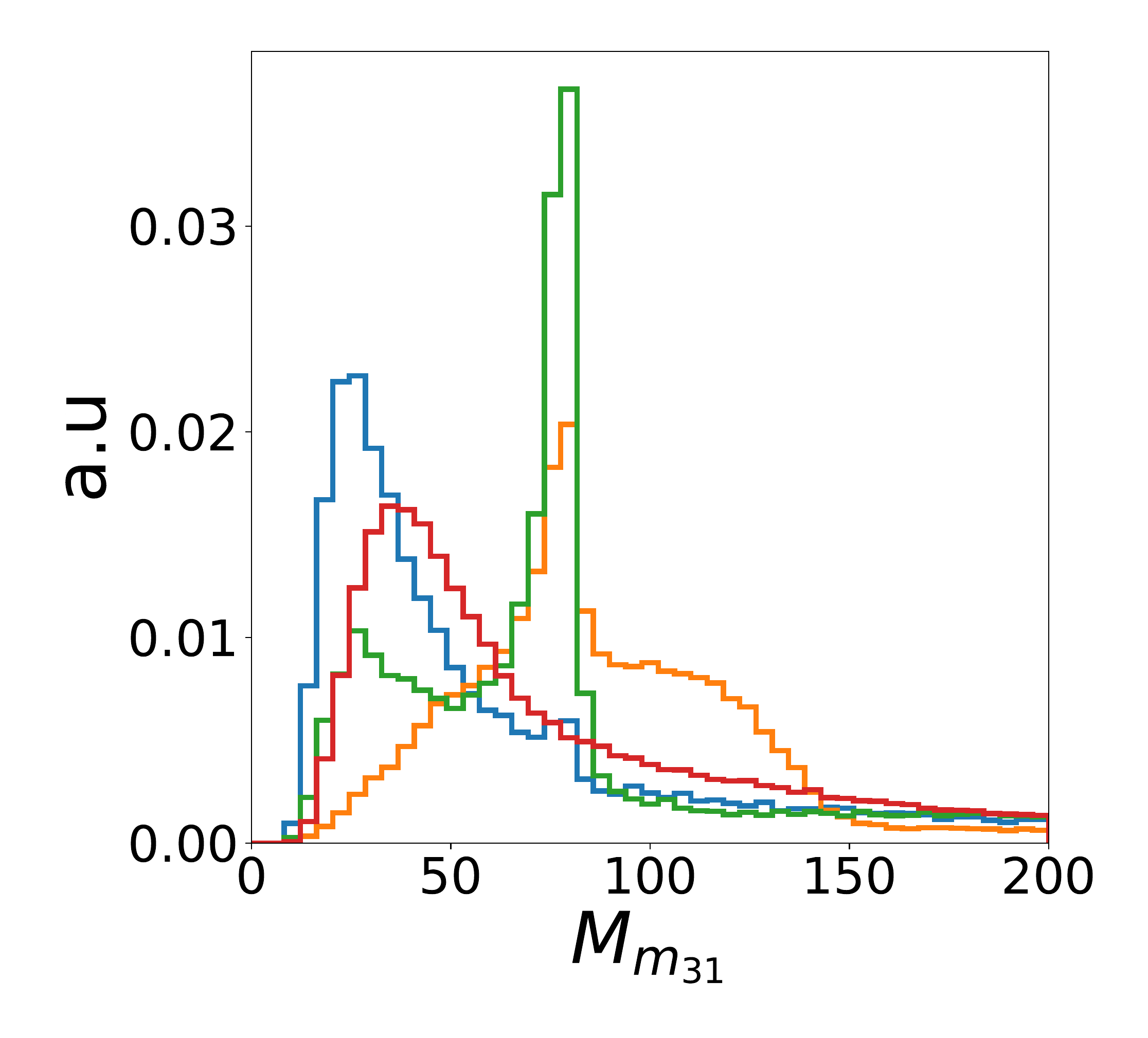}
	\caption{Similar to Fig.~\ref{fig:microjetsR}, but showing the normalised invariant mass distribution between three leading microjets in the fat jet.}
	\label{fig:microjetsM}
\end{figure*}

To map out an infrared safe input to the graph network, we first use the anti-$k_T$ jet algorithm to re-cluster the fat jet constituents into microjets\footnote{As shown in Ref.~\cite{ATLAS:2014twa}, such objects are under good experimental control.}  with a finer resolution of $R=0.1$ and minimum $p_T = 5~\text{GeV}$. We consider fat jets with at least three microjets for our neural network analysis.
 
Identifying each microjet as a node in the network, we construct a graph associated with each jet as follows:

\begin{itemize}
	\item \textbf{Node feature vectors}: We associate five microjet observables as the node's feature vector $\vec{x}$. These are  $\log p_t,\;\Delta \eta\;,  \Delta \phi,\; \Delta R,$ and $\bar{m}$. Here, $p_t$ is the transverse momentum of the microjet, $\Delta \eta,  \Delta \phi,$ and $\Delta R$  are differences in pseudorapidity, azimuthal angle, and angular distance between the microjet and the jet axis respectively. $\bar{m}$ is the mass of the microjet divided by 100 GeV, which, along with the log on $p_t$, reduces the disparity in the range with the other three angular variables.
		\item \textbf{Edge feature vectors}:  After the nodes are defined, we define the graph as the complete graph with all possible edge connections. For each edge, we construct an associated edge-feature vector of three dimensions. Its components are the two distance parameters between the nodes as defined below, and one invariant mass parameter: $\vec{e}_{ij}\equiv(d_{ij}^{\text{CA}},\log d_{ij}^{k_t},  \log m_{ij})$.  The metric $d_{ij}$ is given by 
	\begin{equation*}
	d_{ij} = \min(p_{ti}^{2p},p_{tj}^{2p})\frac{R_{ij}^2}{R^2},
	\end{equation*}
	where $p = 0$ for Cambridge-Aachen (CA) jets, $p = 1$ for $k_t$ jets and $R$ is radius parameter for the fat jet. The CA measure provides information about the geometric distance between two microjets, whereas the $k_t$ measure is motivated from QCD splittings~\cite{Catani:1991hj,Ellis:1993tq}. $m_{ij}$ is the invariant mass of the two microjets. These three variables capture the essential physics between two nodes. 	
		\item \textbf{Adjacency Matrix}: We also construct the adjacency matrix for each edge feature to facilitate their reconstruction at the decoder side. It is defined as
	\begin{equation*}
	\label{eq:adj_mat}
	A^a_{ij}=A^a_{ji}= \left\{\begin{array}{cc}
	e^a_{ij}\quad & \text{if } i \neq j\\
	1\quad & \text{otherwise} 
	\end{array}\right.\quad,
	\end{equation*}
	where $a$ is the vectorial index. Thus, for a jet-graph of $N$ nodes, we have three $N\times N$ matrices.
	The network outputs the edge-features in this representation, and hence the edge-loss is defined as a function of these adjacency matrices.

\end{itemize}
The distribution of $\Delta{R}_{ij}$ and $m_{ij}$ for the 3-leading microjets of each jet are shown in Figs.~\ref{fig:microjetsR} and \ref{fig:microjetsM}. The construction of the graphs and the network analysis are performed using the {\tt Deep Graph Library}~\cite{wang2020deep} with the {\tt PyTorch}~\cite{paszke2019pytorch} backend. 
\begin{figure*}[t]
	\centering
	\includegraphics[width=\textwidth,height=0.45\textwidth]{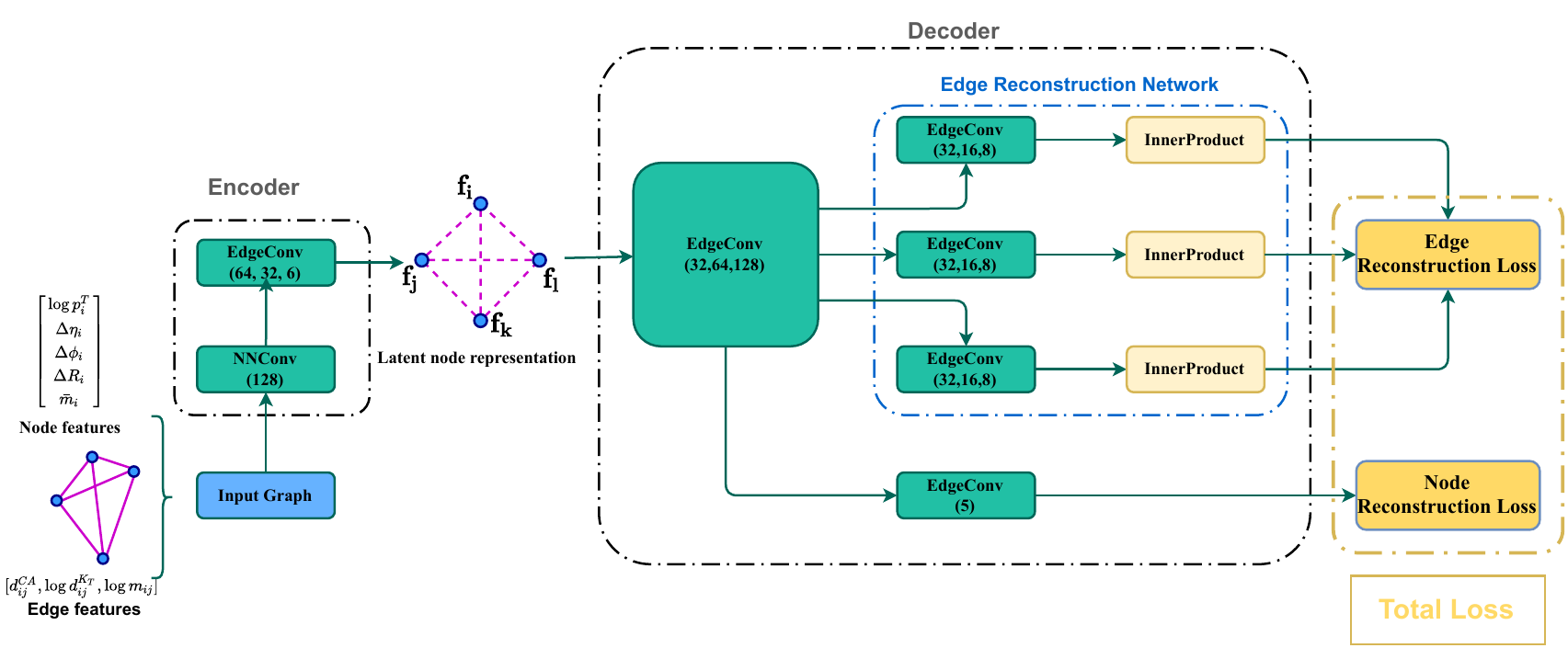}
	\caption{A schematic representation of a graph-autoencoder network. The network contains the (a) Encoder and the (b) Decoder. We employ an edge reconstruction network in the decoder to reconstruct the multidimensional edge information.}
	\label{fig:autoencoder}
\end{figure*}

\section{Graph Neural Networks}
\label{sec:net_arch}
In this section, we describe the various components of our neural network analysis. We briefly detail the conceptual structure of GNNs before moving on to describe the ones we utilise in our analysis, along with the explicit form of the autoencoder's loss function. The network architecture and the process of training are described thereafter.
 
 \bigskip
   
Graph Neural Networks are models that can extract features from graph-structured data. They generalise the inbuilt inductive biases in Convolutional Neural Networks~(CNNs) like local connectivity and shared weights to variable length and possibly non-Euclidean data~\cite{726791}. For supervised learning applications, this was formalised as Message Passing Neural Networks~(MPNNs) in Ref.~\cite{gilmer2017neural}. We sketch the general paradigm and then describe in greater detail the two specific forms that are used in our work in the succeeding paragraphs. In the following, $\vec{h}^{(l)}_{i}$ is the $i^{th}$ node's features at the $l^{th}$ timestep (analogous to a layer in the usual ANNs). $\vec{e}^{(l)}_{ij}$ denotes the features of the edge connecting the nodes $i$ and $j$, and $\mathcal{N}(i)$ is the set of nodes connected to the node $i$. For the input layer, we take $l=0$, and $\vec{h}^{(0)}_i=\vec{x}_i$. MPNNs consist of a message passing phase and a  graph readout layer. In the first phase, a message-passing function is defined for two nodes $i$ and $j$
\begin{equation}
  \label{eq:msg_pass}
  \vec{m}^{(l)}_{ij}= \vec{M}^{(l)}(\vec{h}^{(l)}_i,\vec{h}^{(l)}_j,\vec{e}^{(l)}_{ij})\,,
  \end{equation}
  which calculates the message $\vec{m}_{ij}$ for the edge connecting the nodes.
  The message function is usually a multilayer-perceptron~(MLP) shared between all the edges, hence the term graph convolutions. For each timestep (or layer), the messages between all connected nodes are calculated, after which the features of each node are updated according to an aggregation function
  \begin{equation}
  \label{eq:agg}
  \vec{h}^{(l+1)}_i=\square(\vec{h}^{(l)}_i,\{\vec{m}^{(l)}_{ij}|\, j\in\mathcal{N}(i)\})\,,
  \end{equation}
  which is a function of the node feature and all incoming\footnote{The message passing function $M$ need not be symmetric in $\vec{h}_i$ and $\vec{h}_j$.} messages. The updated features $\{\vec{h}^{(l+1)}_i\}$, are used for the next timestep to perform another message passing and aggregation step. At the output of the final message-passing timestep, a graph readout layer performs a permutation-invariant\footnote{A permutation invariant function makes the output invariant to graph isomorphisms.} operation (for instance, max, sum, or mean) on the node features to give fixed-length vectors, regardless of the number of nodes. In supervised learning, this vector can be the final output which can be used either to minimise the loss function or fed into an MLP. However, in a graph-autoencoder,  graph-level readouts are not applied to preserve the graph structure until the final output. 
  Graph-autoencoders are typically designed for classifying nodes or edges, focusing on learning local features of a huge graph. However, as our goal is to classify small graphs, the network needs to learn global graph structures {\emph{and}} local features. To overcome this, we design an edge-reconstruction network within the decoder, making our network capable of learning graph structures by reconstructing the graph in its entirety. The complete structure of our network is shown in Fig.~\ref{fig:autoencoder}, where the black boxes encase the encoder and the decoder. The edge-reconstruction network is shown bounded by the blue box. These are described in greater detail in the following passages.

\subsection{Autoencoder}

Autoencoders are neural networks that map an input space to a bottleneck dimension (the latent dimension) and then back again to a space identical to the input. We use the graph-convolutions proposed in Ref.~\cite{gilmer2017neural} to incorporate the multi-dimensional edge information along with the input node features. Our network, therefore, learns the physics information that is encoded into our 3-dimensional edge feature. The timesteps until we reach the latent space employ edge-convolution~\cite{wang2019dynamic}, which has proved excellent performance in supervised learning scenarios~\cite{Qu:2019gqs,Dreyer:2020brq}. We refer to these two layers as \emph{NNConv}, and \emph{EdgeConv} respectively, according to the python class name implemented in the {\tt Deep Graph Library}. The encoder block outputs a graph with the same edge connections as that of the input with updated latent features $\vec{f}_i$ for each node. The decoder reconstructs the node and edge features from this latent node representation. As shown in Fig.~\ref{fig:autoencoder}, the decoder has a shared block of edge convolutions, after which the output feeds into four different blocks of edge convolutions: a single layer for the node reconstruction, and three edge reconstruction blocks. These three blocks are identical in structure and reconstruct each edge feature independently from the propagated information from the shared block. We use an \emph{Inner Product Layer}~\cite{kipf2016variational} to reconstruct the edge information in the form of three adjacency matrices. These three components and the composition of the loss function are explained in the subsequent paragraphs.

\medskip
\noindent {\bf NNConv:} 
The first layer takes the node and edge features as input and performs a weighted graph convolution by making use of an MLP, referred to as edge function $F_w$. This takes the edge features as input and maps it to a dimension of $m\times n$, where $m$ is the input node's dimensions (5 in the present case), and $n$ is the dimension of the updated node features. The message passing function performs a broadcasted element-wise multiplication of the form 
\begin{equation}
\label{eq:nn_broad}
{}^{ab}m^{(1)}_{ij}={}^{ab}F_e (\vec{e}_{ij}) \times \; {}^{ab}\tilde{h}_j^{(0)}\,,
\end{equation} 
where $a$ and $b$ are the indices of the matrix, and ${}^{ab}\tilde{h}_j^{(0)}$ is formed by repeating $\vec{h}^{(0)}_j$, the input node features, $n$ times. The aggregation step takes the mean of ${}^{ab}m^{(1)}_{ij}$ over all neighbouring nodes $j$, and then sums over the $a$ index of the matrix:
\begin{equation}
\label{eq:nnconv}
{}^{b}h_{i}^{(1)} = \sum_a \mathrm{mean}_{j\in \mathcal{N}(i)}\left(\left\{{}^{ab}m^{(1)}_{ij}
 \right\}\right)\,,
\end{equation}
to give updated $n$ dimensional node features $\vec{h}_{i}^{(1)}$. 

\medskip
\noindent {\bf EdgeConv:} The backbone of our architecture is the edge convolution operation~\cite{wang2019dynamic}.  This involves two linear layers: $\Theta_w$ and $\Phi_w$, with identical input and output dimensions, which determine the dimensions of original and updated node features respectively. The message passing function is defined as
\begin{equation}
\label{eq:edge_conv}
\vec{m}_{ij}^{(l)} = \Theta_w (\vec{h}_j^{(l)} - \vec{h}_i^{(l)}) + \Phi_w( \vec{h}_i^{(l)})\,,
\end{equation}
while the aggregation step involves taking the maximum value
\begin{equation}
\label{eq:edge_agg}
{}^ah_{i}^{(l+1)} = \max_{j \in \mathcal{N}(i)} \{{}^am^{(l)}_{ij}\}\,,
\end{equation}in each component $a$ of the incoming message vectors to give the updated node features $\vec{h}^{(l+1)}_i$.

\medskip
\noindent {\bf Inner Product Layer:} The edge-reconstruction network uses an Inner Product Layer to reconstruct the edge features from the node features of the final edge convolution output. The inner product makes the correspondence to the two-node indices for each edge. Since our graphs are undirected, the layer constructs a symmetric $N\times N$ matrix, $N$ being the number of nodes in the graph. Its components are therefore 
\begin{equation}
\label{eq:inn_prod}
\hat{A}_{ij}=\vec{h}_i\;.\;\vec{h}_j\,,
\end{equation}
where $\vec{h}_i$ and $\vec{h}_j$ are node-feature vectors.

\medskip

\noindent {\bf Loss Function:}
We use root-mean squared error (RMSE) for the node as well as the edge reconstruction losses. For the node feature this is
\begin{equation}
\label{eq:node_loss}
L_{node}=\sqrt{\sum_{ia}\frac{(\hat{x}^a_{i}-x^a_{i})^2}{N\times 5}}\,,
\end{equation}where $a$ is the node-feature index, $i$ is the node index, $\hat{x}^a_i$ and $x^a_i$ are the reconstructed and input node features, respectively. We define the edge reconstruction loss as the sum of three individual RMSEs for each edge feature\begin{equation}
\label{eq:edge_loss} 
L_{edge}=\sum_{a}\sqrt{\sum_{ij}\frac{(\hat{A}^a_{ij}-A^a_{ij})^2}{N\times N}}\quad,
\end{equation}where $a$ is the edge-feature index, $i$ and $j$ are node indices. $\hat{A}^a_{ij}$ and $A^a_{ij}$ are the reconstructed and input adjacency matrices respectively. The total loss is the weighted sum of the individual losses,
\begin{equation}
	\label{eq:loss_auto}
L_{auto}=\lambda_{node}\;L_{node}+\lambda_{edge}\;L_{edge}
\end{equation} We choose $\lambda_{node}=0.3$ and $\lambda_{edge}=1$, so that the combined node features get the same weight as each individual edge feature, which carry more relevant physics information. Note that the loss function is invariant to node permutations of the input graph since, mean is a permutation invariant function, and the architecture respects permutation invariance: any change in the node ordering changes the output of each layer(via the graph readout) in conjunction with the adjacency matrix. Our network however, does not reconstruct an arbitrarily permuted graph for a given input, which is not strictly necessary since we concentrate on the reconstruction error of a single graph and not of an equivalence class of graphs.

\subsection{Network Architecture and training}
Neural networks require a careful optimal choice of hyperparameters. As this is a proof-of-principle analysis, we do not perform an extensive hyperparameter scan. However, we scan over the latent dimension, which is critical for any autoencoder. For the first layer of the graph-encoder (NNConv), we use an MLP of hidden dimensions: 256, 128, 64, and 32 as the edge function to map the 3-dimensional edge features to a $5\times 128$ dimensional output. The hidden layers have ReLU activations, whereas the final layer has a sigmoid activation. The limited range of the sigmoid activation helps in giving the addition operation in aggregation (as defined in Eq.~\eqref{eq:nnconv}) an interpretation of a weighted sum over messages in an additional dimension without the dynamics being entirely dominated by the outputs of the edge function. Each hidden layer has a dropout layer with fraction 0.2 of disconnected nodes between layers to avoid overfitting and achieve better generalisation. After the aggregation, we get a 128-dimensional output that feeds into a series of edge-convolution layers with linear layers as $\Theta_w$ and $\Phi_w$. The output dimensions of the linear layers are 64 and 32 and outputs a 6 dimensional latent node encoding. This value is chosen after a scan over different latent dimensions which we elaborate on in the next section. The shared block of the decoder uses the encoder's reversed dimensions: 32, 64, and 128. With the 128-dimensional vector as input, the node reconstruction layer performs an edge-convolution to give the reconstructed node vectors $\hat{\vec{x}}$. Similarly, each edge reconstruction network has three successive edge convolutions of output dimensions 32, 16, and 8. We calculate the inner products on the 8-dimensional vector space to give the reconstructed adjacency matrices $\hat{A}^a_{ij}$. 

We train the network with the Adam optimiser~\cite{kingma2014adam} initialised with a 0.001 learning rate on mini-batches of 64 samples. The learning rate is decayed with a reduce-on-plateau condition with decay factor 0.5, and a patience of five epochs with an additional five epochs of cool-down. We use 85k jets to train the network.  After each epoch, we calculate the loss of an independent validation dataset containing 28k QCD jets. We stop the training once the learning rate goes below $10^{-8}$. The epoch with minimum validation loss is used for further inference.

\section{Results and Discussion}
\label{sec:results}
\begin{figure*}[t!]
	\centering
	\subfigure[\label{fig:auc_lat}]{\includegraphics[scale=0.24]{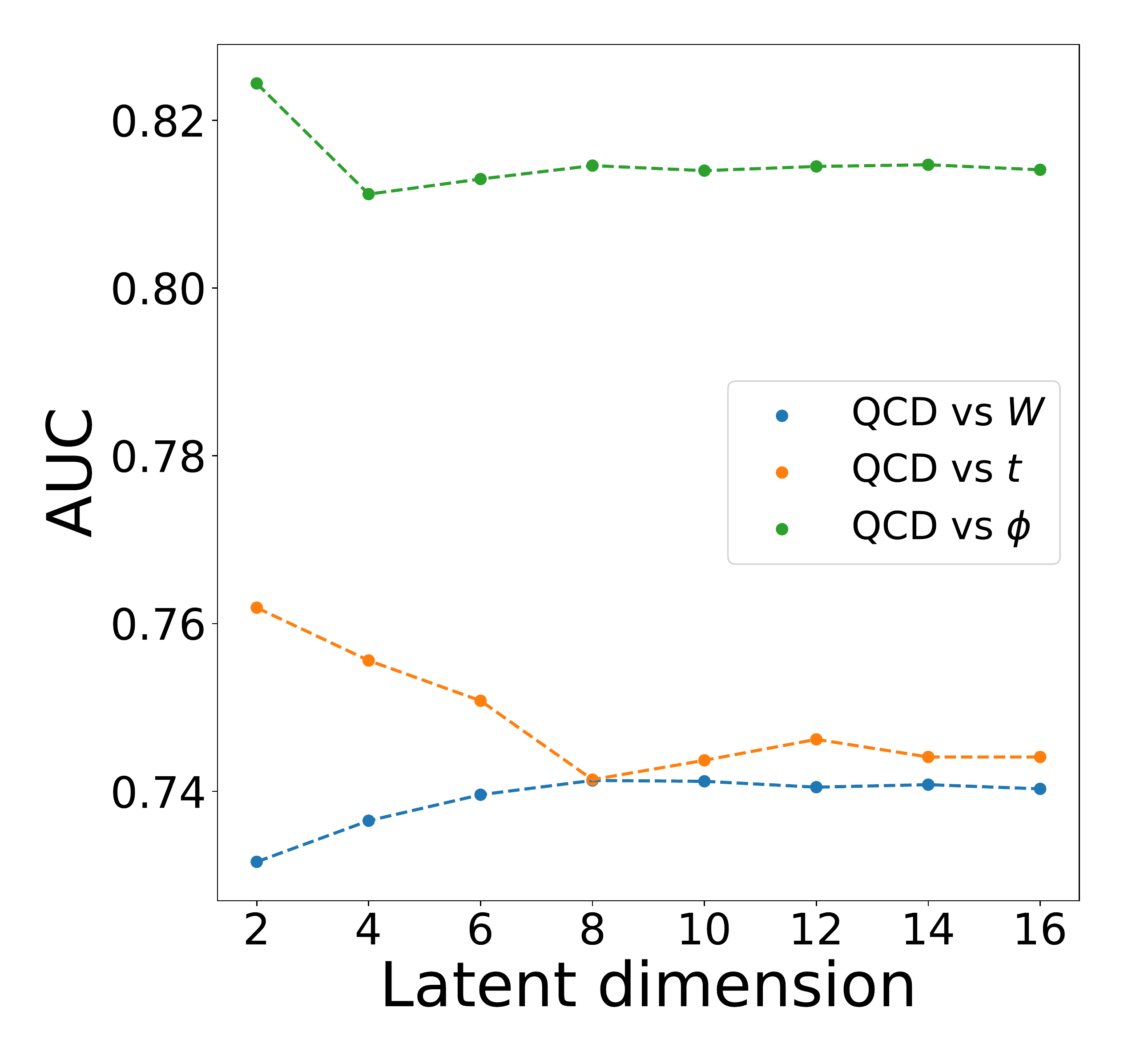}}
	\hskip 0.4cm
	\subfigure[\label{fig:mean_loss}]{\includegraphics[scale=0.24]{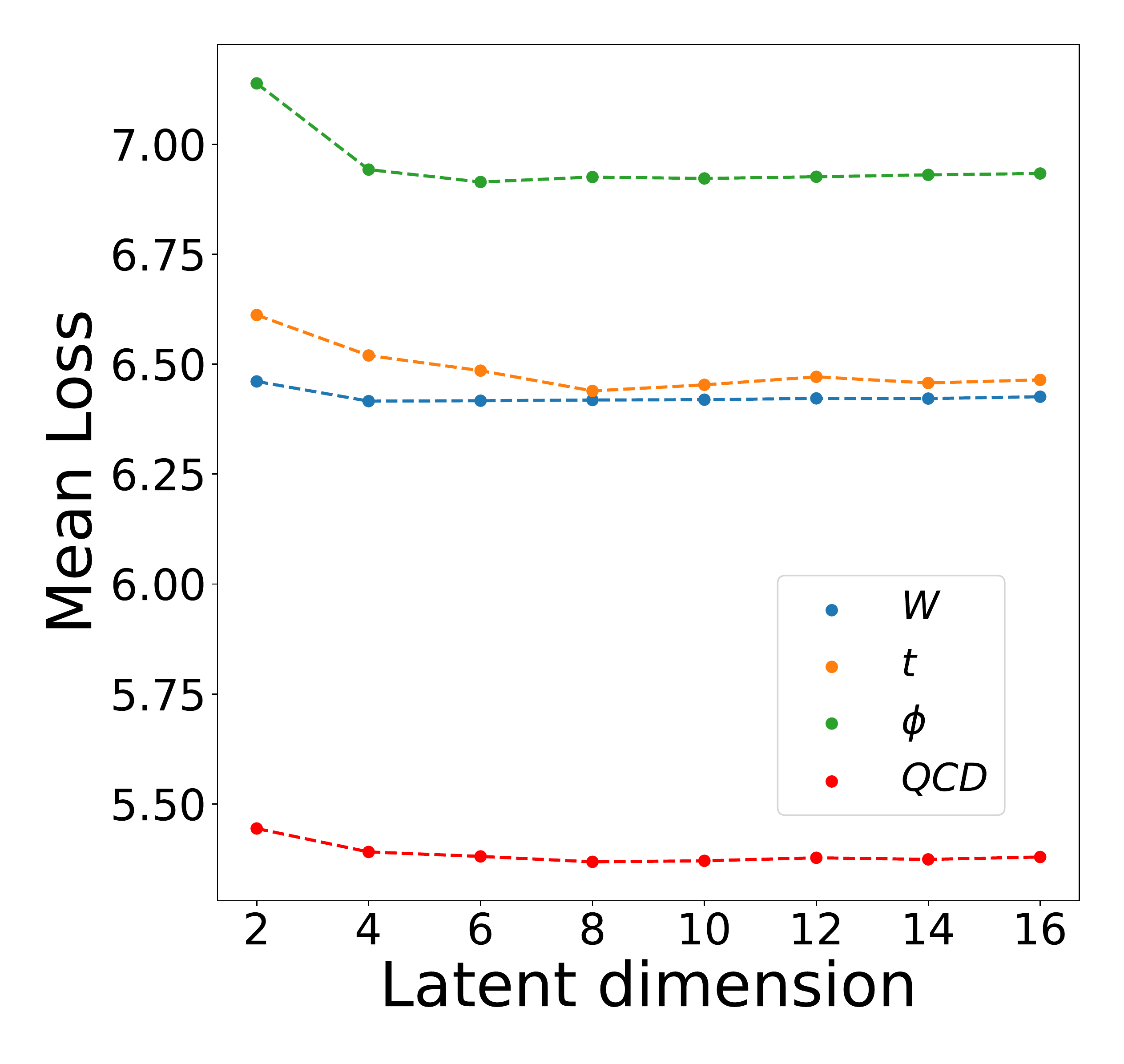}}
	\caption{The AUC and mean loss for the three signal classes as a function of latent dimension from 2 to 12 for the given architecture}
	\label{fig:AUC_mean}
\end{figure*}
\begin{figure*}[th!]
	\centering
	\subfigure[\label{fig:loss_dist}]{\includegraphics[scale=0.25]{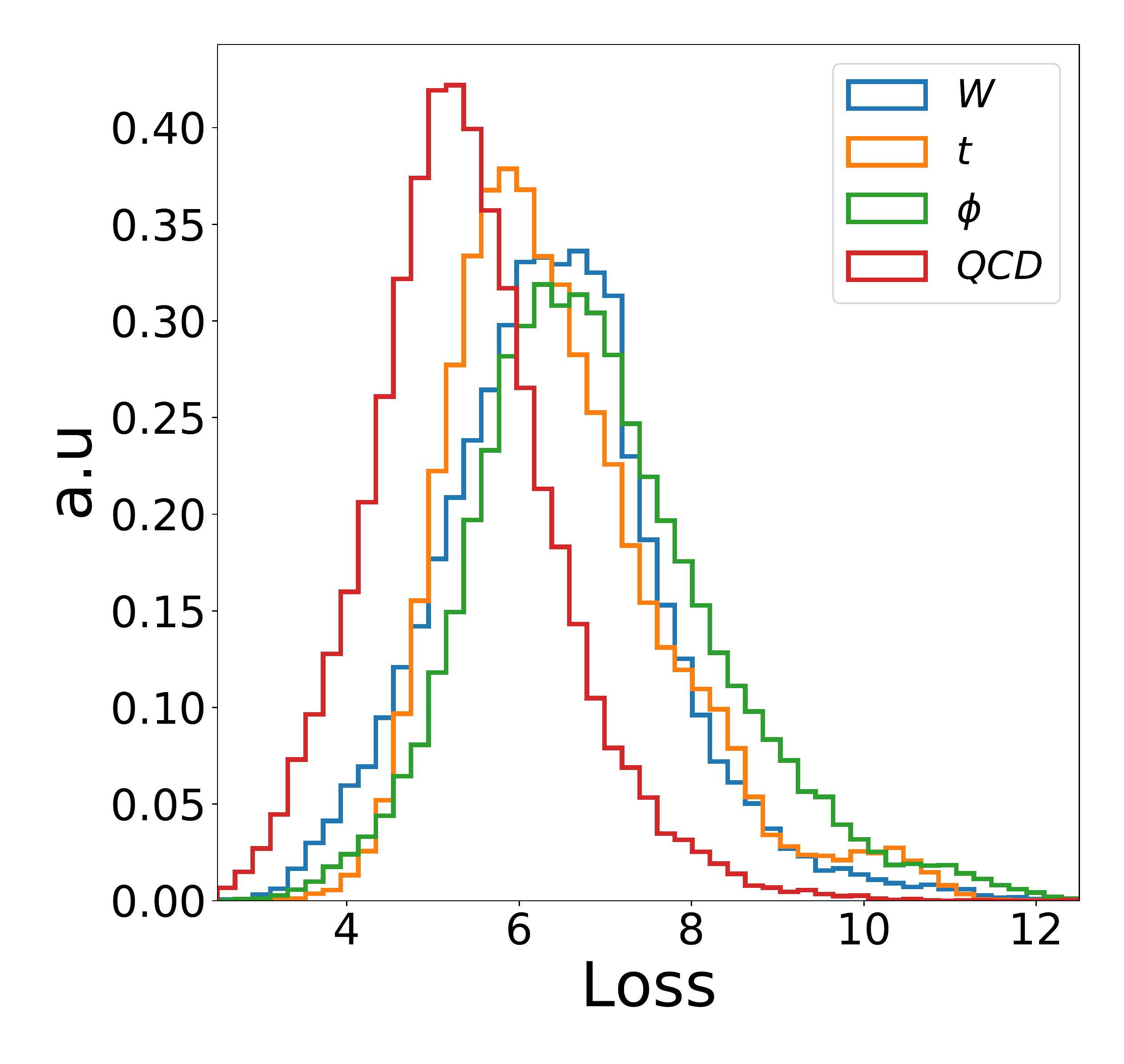}}
	\hskip 0.4cm
	\subfigure[\label{fig:loss_roc}]{\includegraphics[scale=0.25]{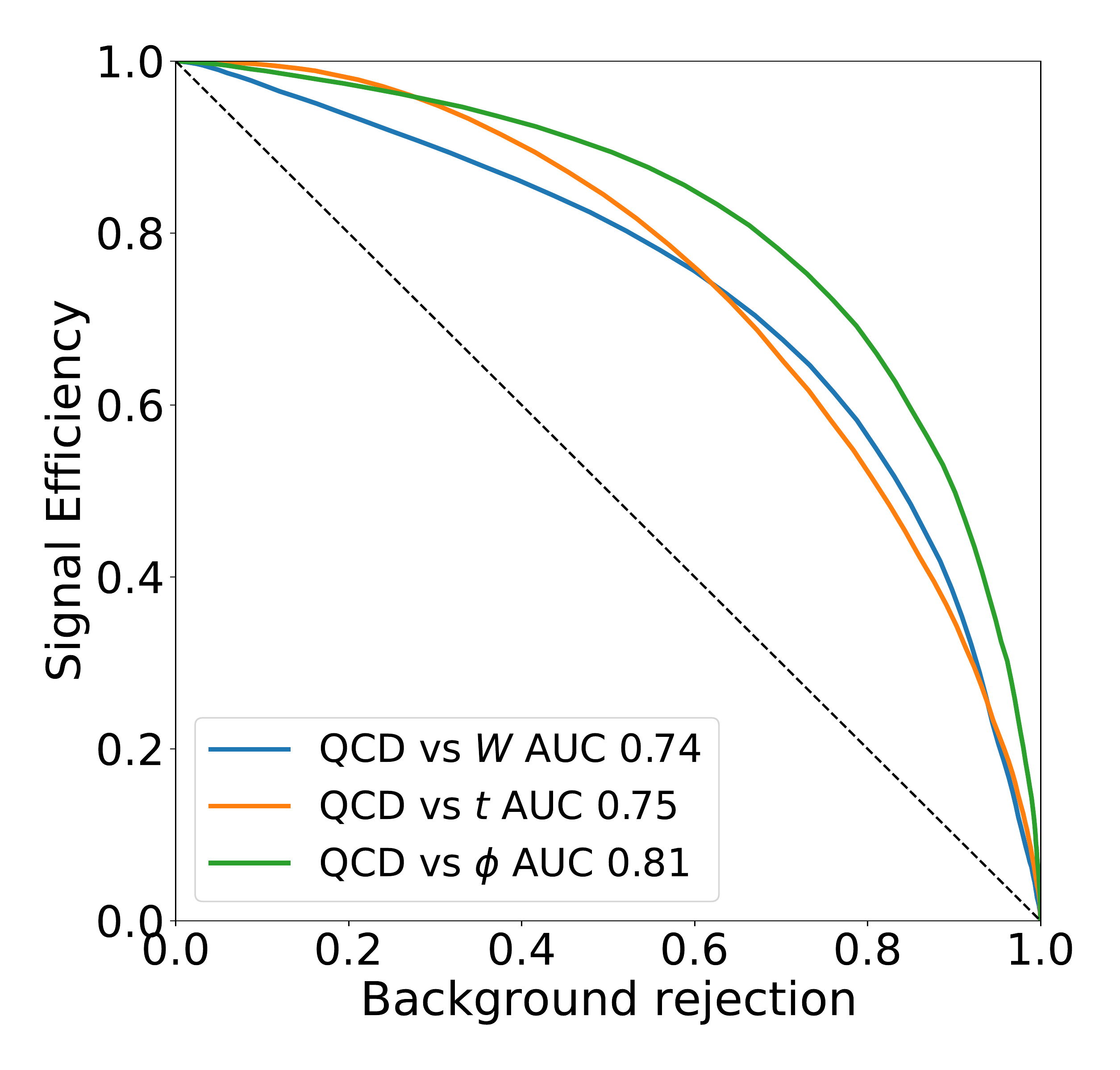}}	
	\caption{The loss of the graph-autoencoder (a) and ROC curves (b) for a network trained only on QCD jets.}
	\label{fig:loss_dist_auc}
\end{figure*}
\begin{figure*}[t]
	\centering
	\includegraphics[scale=0.23]{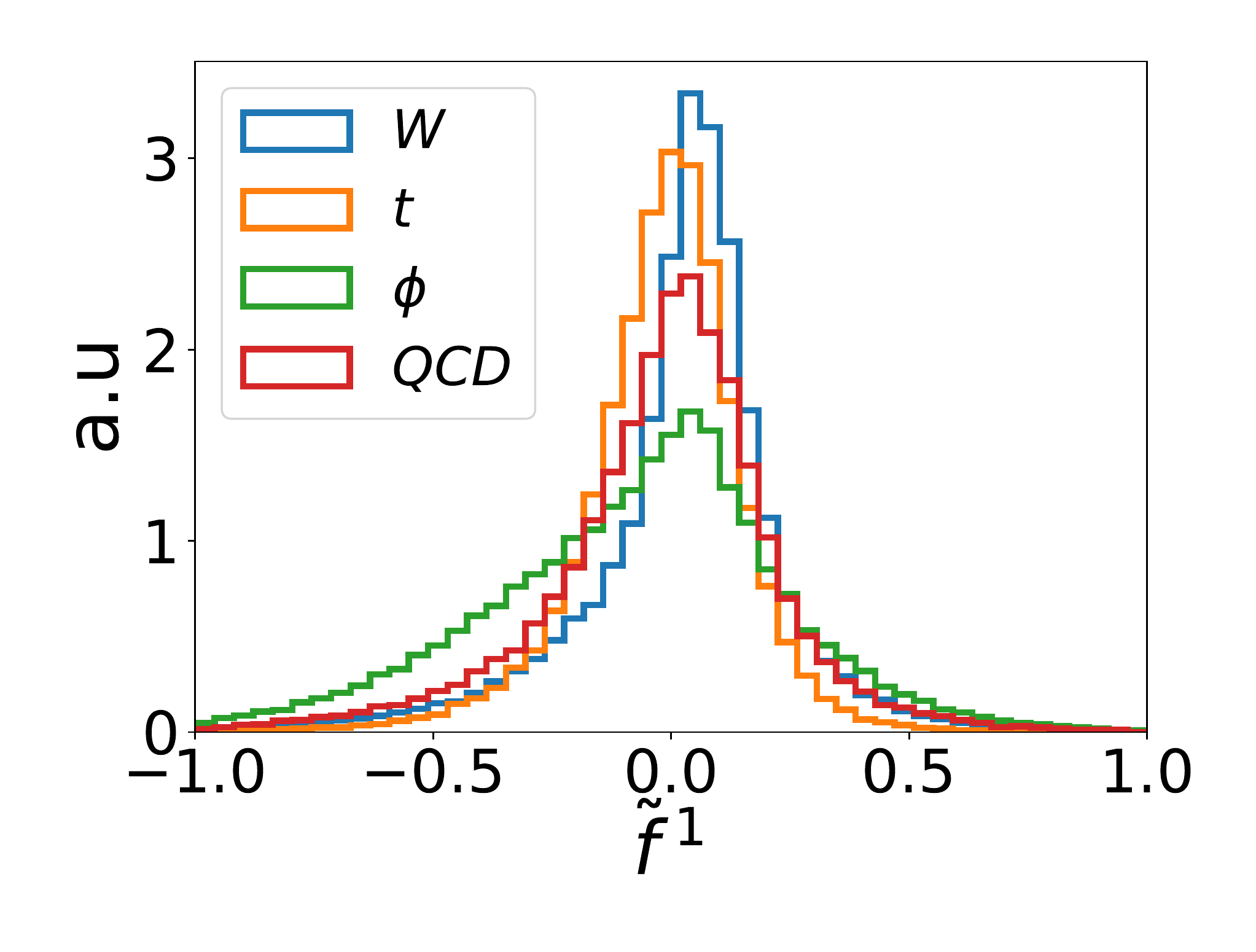}
	\includegraphics[scale=0.23]{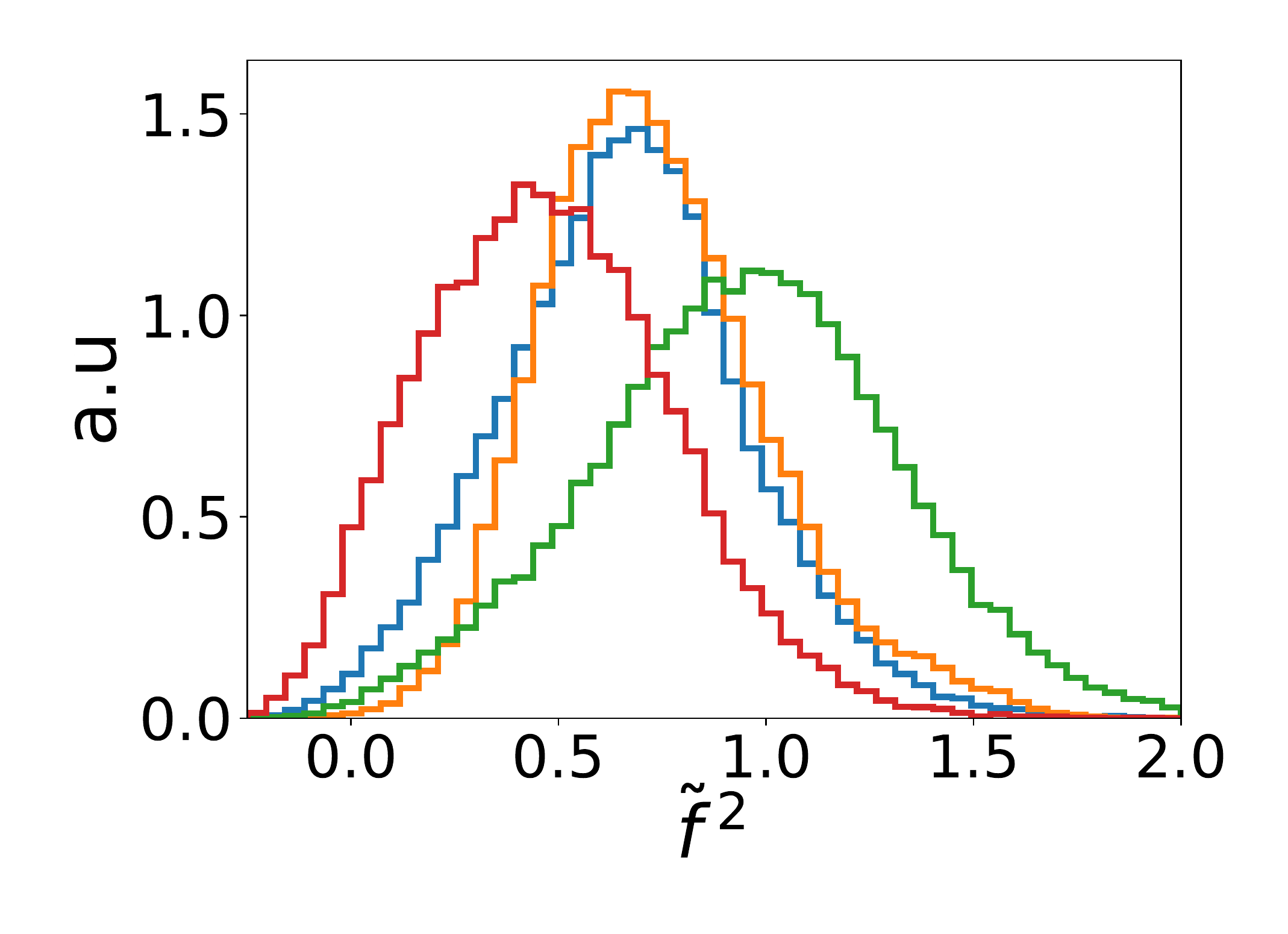}
	\includegraphics[scale=0.23]{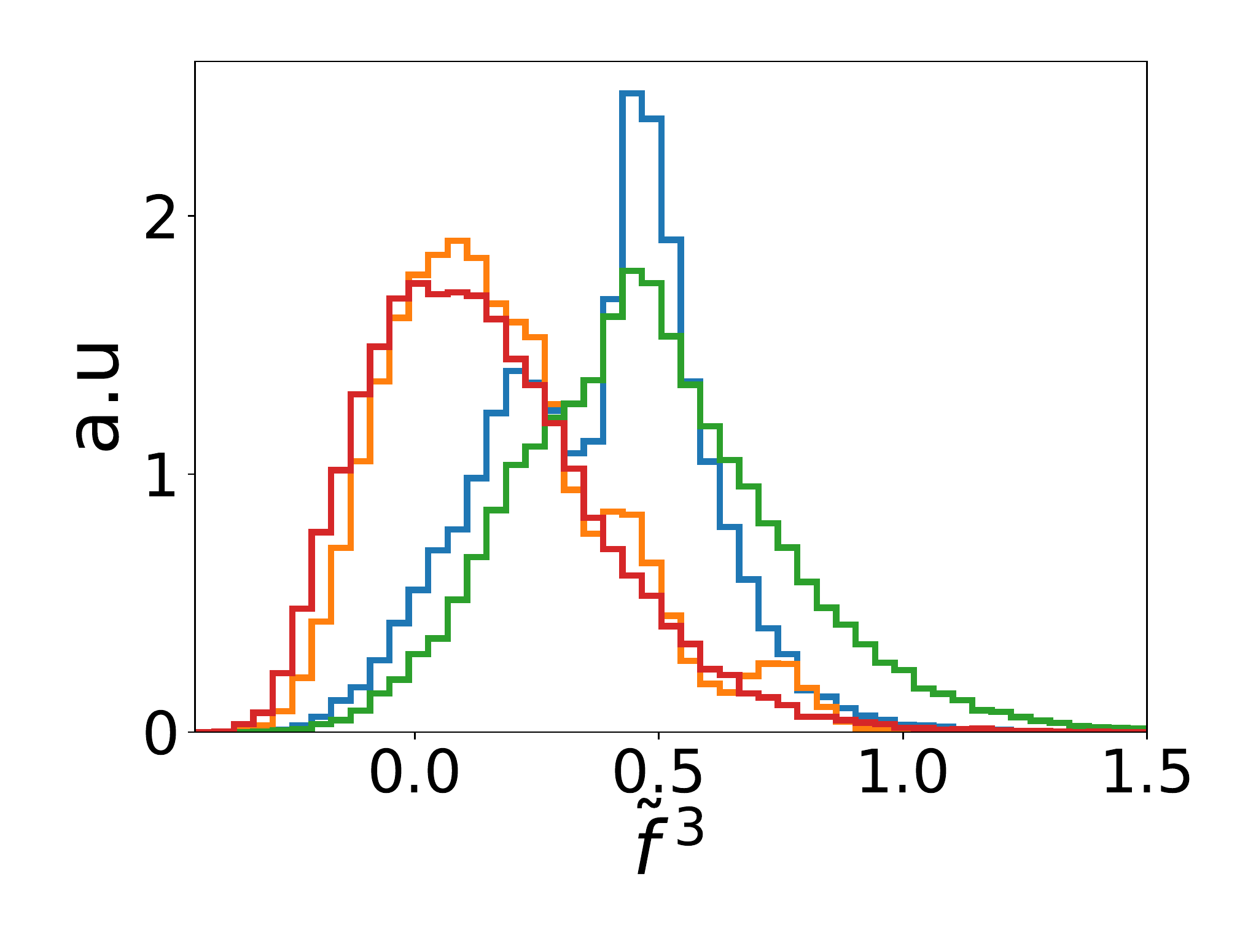}\\
	\includegraphics[scale=0.23]{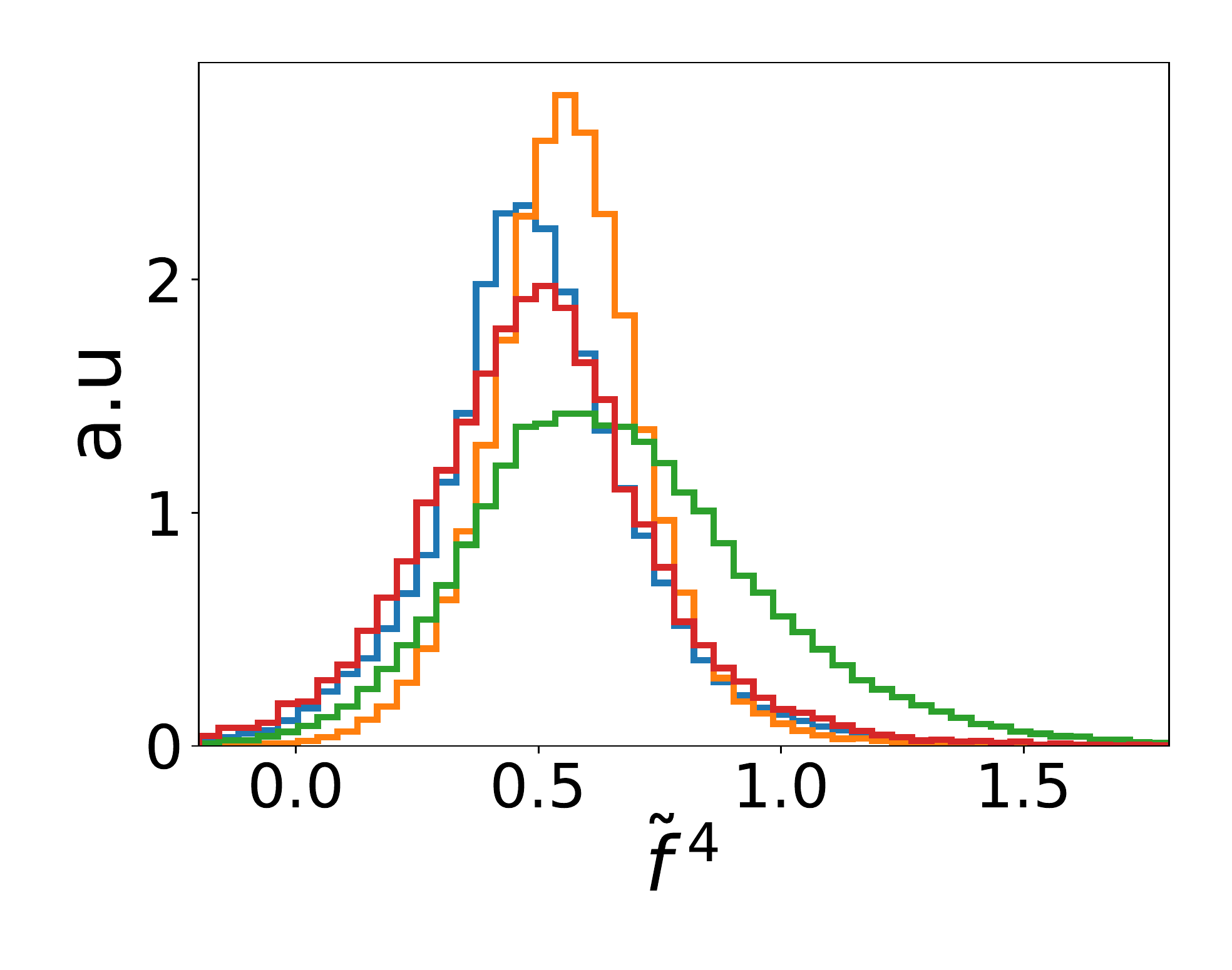}
	\includegraphics[scale=0.23]{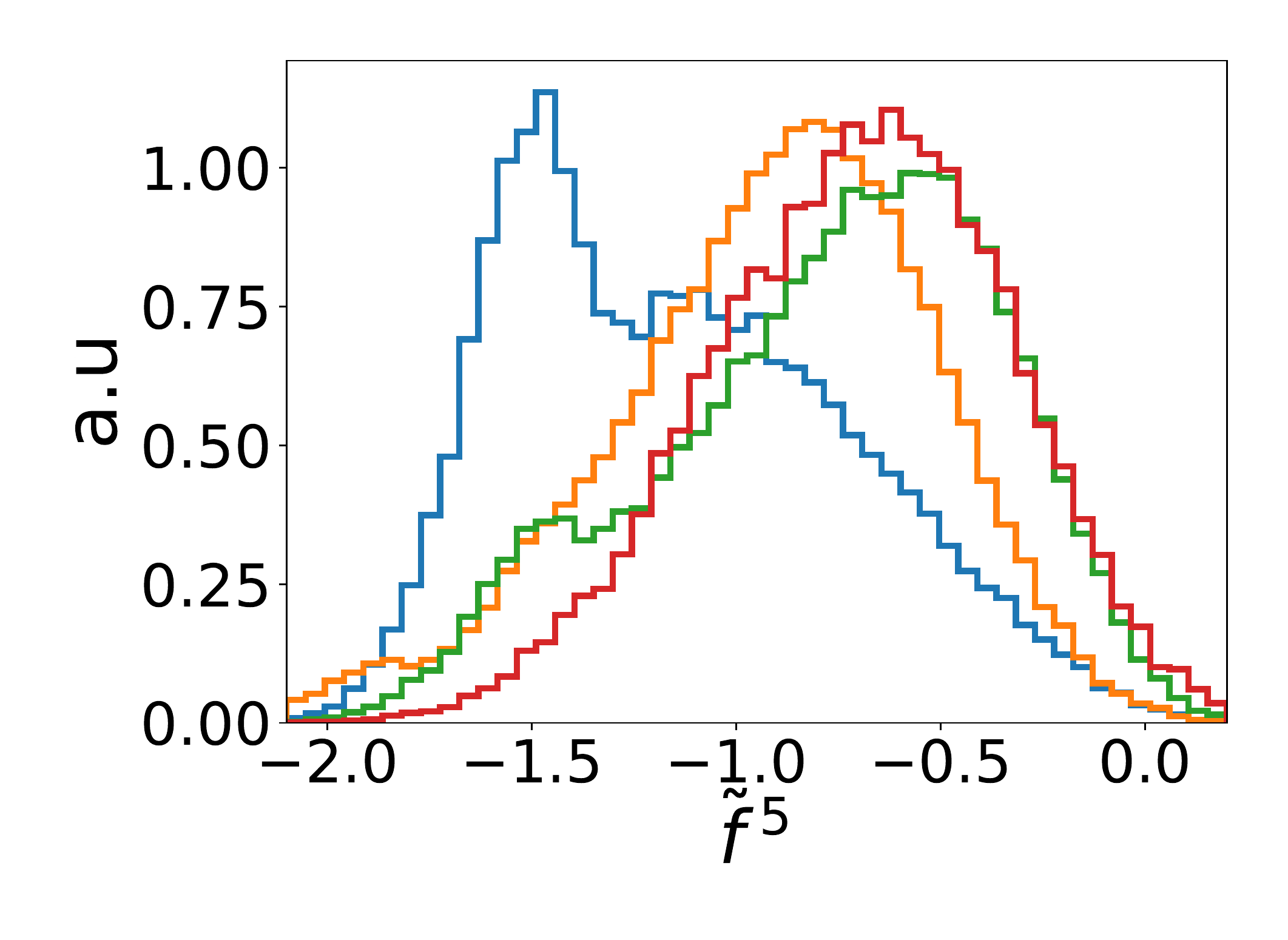}
	\includegraphics[scale=0.23]{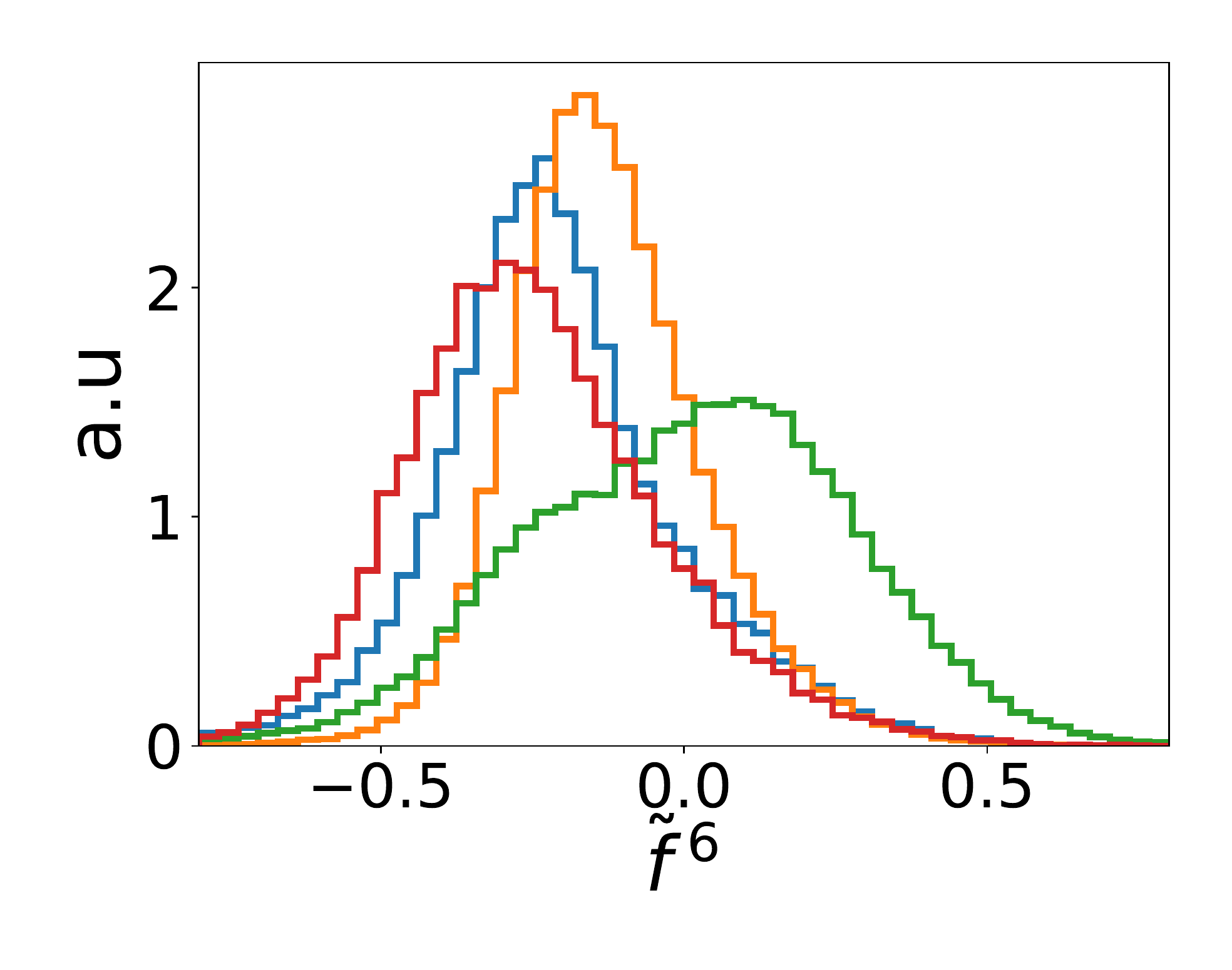}	
	\caption{The distribution of six dimensional latent space after the training is performed only on QCD jets.}
	\label{fig:latent_dist}
\end{figure*}
\begin{figure*}[!t]
	\centering
	\includegraphics[scale=0.135]{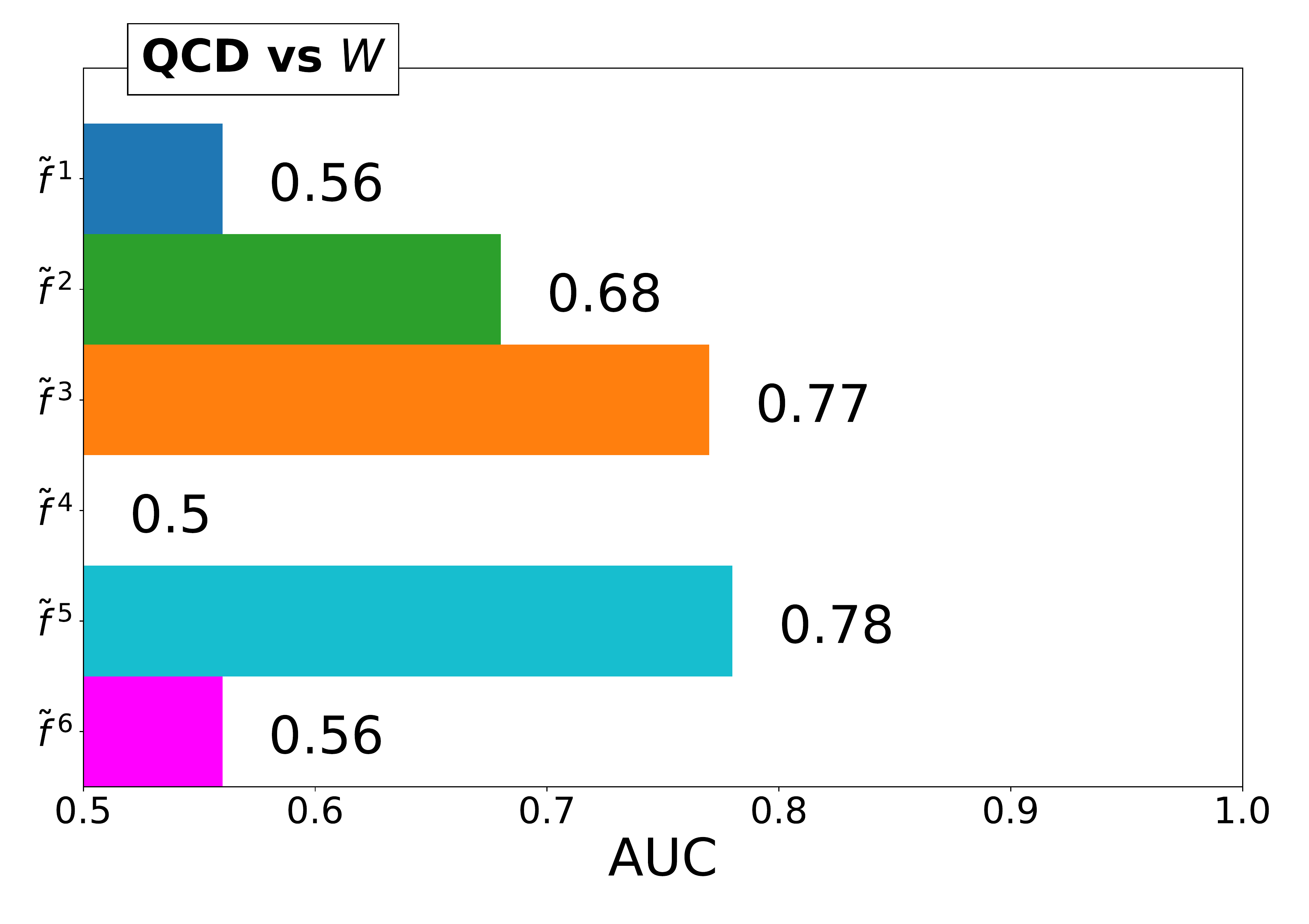}
	\includegraphics[scale=0.135]{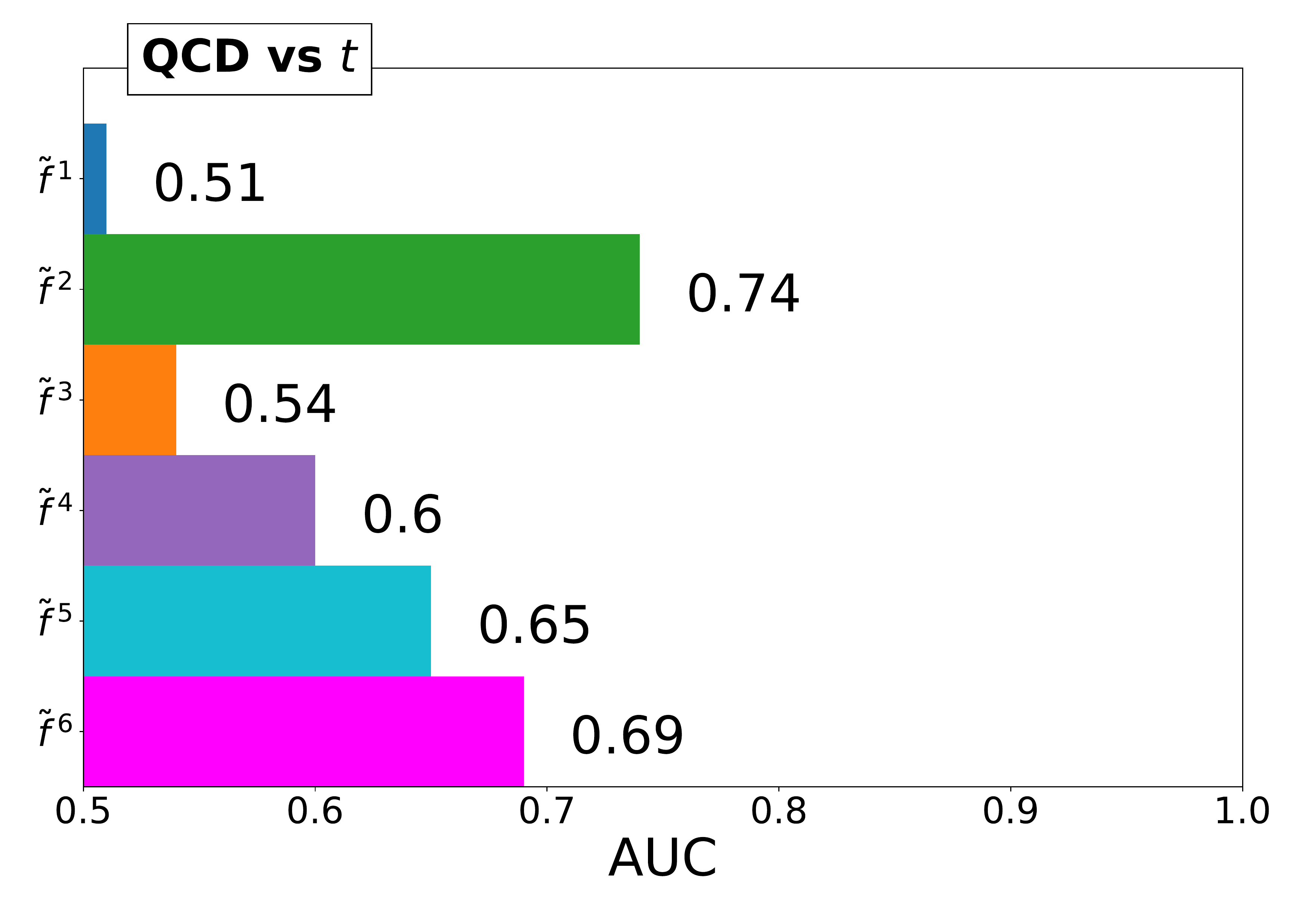}
	\includegraphics[scale=0.135]{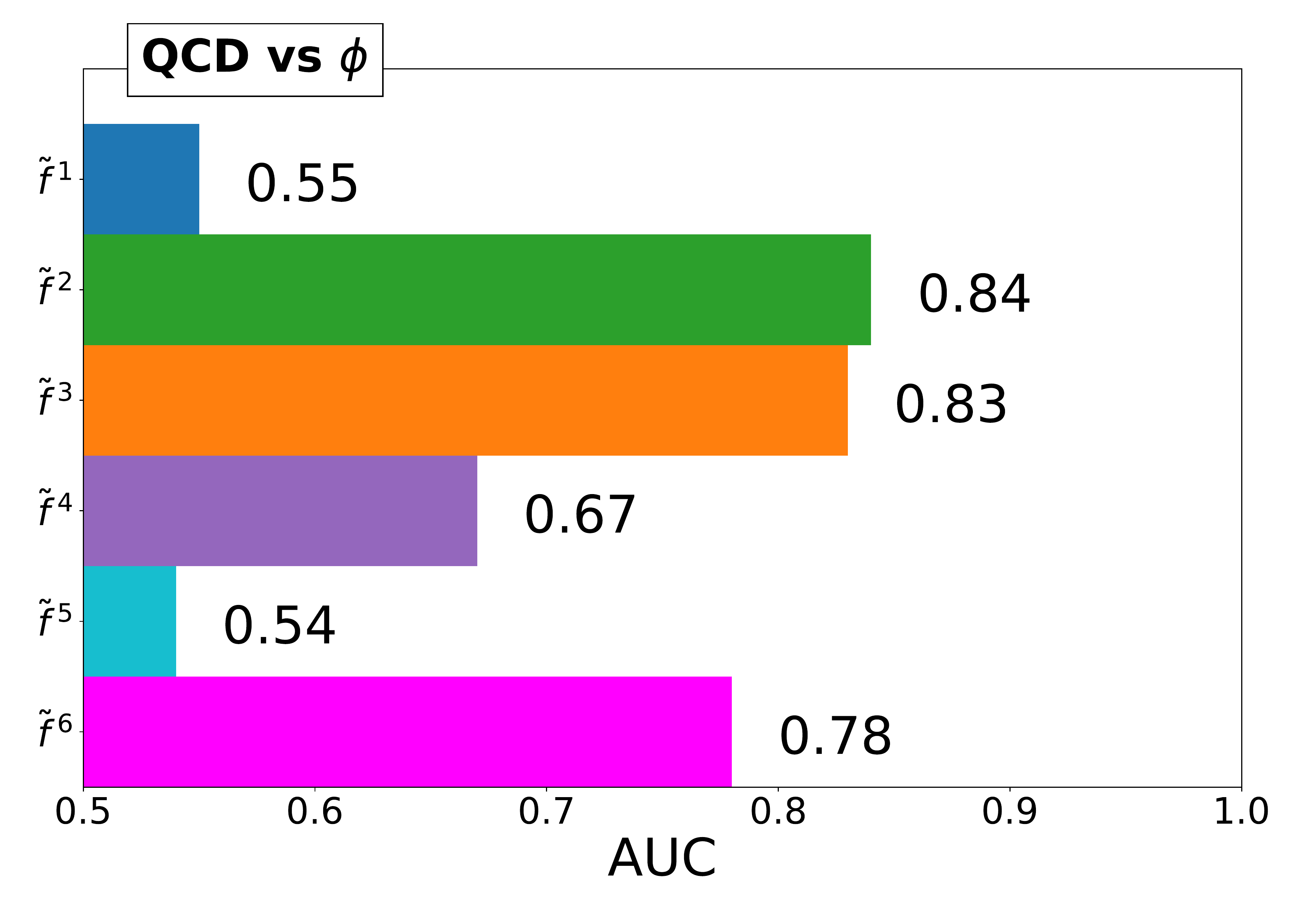}
	
	\caption{The AUC for all three signal classes corresponding to each latent dimension in the network.}
	\label{fig:latent_AUC}
\end{figure*}
In order to test the performance of the graph-autoencoder for the different non-QCD signals described in Sec.~\ref{sec:details}, we evaluate the discrimination power of the total loss function as defined in Eq.~\eqref{eq:loss_auto}. We use an independent testing data set of 28k for QCD jets and a similar number for the signal samples. We first scan the latent dimension from 2 to 12 in multiples of two, keeping all other hyperparameters fixed. 
The Area-Under-(the)Curve (AUC) between the signal acceptance and the background rejection for each latent dimension is shown in Fig.~\ref{fig:auc_lat}. In Fig.~\ref{fig:mean_loss}, we show the mean loss for each class as a function of the latent dimension. We can see that although the mean loss is relatively stable for QCD jets after 4-dimensions, the AUC of the different signal vs. QCD scenarios varies significantly.  The variation is due to the unsupervised nature of the algorithm; the network has no information about the signal classes.

On the other hand, from the different nature of the AUC curves, we can understand the information passed for differing latent dimensions. The increasing AUCs for the $W$ classification hints that the network sees them as similar to QCD jets when the information passed in the bottleneck is smaller, but the features of a typical QCD jet are not fully modelled for low dimensions, thus making this discrimination not reliable.
Increasing the bottleneck dimension makes the network learn QCD features, which then leads to robust anomaly detection for top quarks and $W$ bosons.
The $\phi$ jets, which have the most noticeably different structure from QCD jets, reach a stable AUC much faster than tops and $W$ bosons.
We infer that latent dimension of $\sim 6$ shows a stable performance for all three classes (in particular for QCD jets) and has reached the plateau in the mean loss. Since we cannot optimise the network to each class in anomaly detection, we fix six as the latent dimension parameter.  
The normalised distribution of the loss function for all classes is shown in Fig.~\ref{fig:loss_dist}.
As the network is trained using QCD jets, the autoencoder reconstructs them with lower loss, while all signal classes have a relatively higher loss. By vetoing the QCD jets with lower losses, we tag (new physics) signal jets (anomalous class); the Receiver-Operator-Characteristic (ROC) curve between the signal acceptance and background rejection is shown in Fig.~\ref{fig:loss_roc}\footnote{We compare the results obtained for our dataset with particle graph autoencoders used in Ref~\cite{kasieczka2021lhc} in Appendix~\ref{app:pgae}.}. 
The performance increases as the prong structure becomes richer for the signal classes. We discuss the correlation of the loss with some important jet-level observables in Appendix \ref{app:loss_corr}.

We also investigate the latent representation learned by the graph-autoencoder to explore compressed representations for QCD jets. Latent representations have also been investigated in similar, and indeed different, physical scenarios recently in Refs.~\cite{Dillon_2020,bortolato2021bump,dillon2021better}. Even though we do not perform graph readouts during the training, the graph-autoencoder learns the graph structure via the edge reconstruction network. We use a graph readout that takes the mean in each dimension of the latent node features to obtain a fixed-dimensional latent graph representation. More precisely, we consider 
\begin{equation*}
\tilde{f}^a=\frac{1}{N}\sum_{i\in G}\;f^a_i\,,
\end{equation*} 
where $a$ is the vector-index, $i$ is the node index and $G$ is the set of all nodes of the graph.  The normalised distribution of the four classes for each latent dimension is shown in Fig.~\ref{fig:latent_dist}, while the corresponding AUC for each signal vs. QCD discrimination is shown in Fig.~\ref{fig:latent_AUC}.  We find that $\tilde{f}^2$ performs best for top and $\phi$ jets,  while $\tilde{f}^5$ gives the maximum AUC for $W$ jets.  The AUC for top quarks and scalar $\phi$ from $\tilde{f}^2$ is 0.74 and 0.84 respectively. Thus, we find a significant improvement for $\phi$ from the value obtained with the loss function, which is also the case for the $W$ jets whose AUC is 0.78 from $\tilde{f}^5$. The latent distributions are prone to training uncertainties since they do not have any regularising terms in the loss function.

	More precisely, the shapes and location of these distributions will vary significantly for different training instances even when they give very similar distributions of the loss function. There are available remedies~\cite{kingma2014autoencoding,makhzani2015adversarial,patrini2019sinkhorn} for the training class, but controlling the signal distributions during unsupervised training is not possible by design. However, it may be possible to control them using physically motivated priors, which is beyond the scope of our present work. Nevertheless, once we have a single training instance, latent dimension-based anomaly finders can be used by trimming the encoder network after training to contain only these two outputs. Control samples can be used to quantify the latent space distributions and could therefore find applications in trigger optimisation.

\section{Conclusions}
\label{sec:conc}
In this work, we have introduced a graph neural network-based autoencoder for unsupervised anomaly detection in QCD boosted jet data.  We design a novel edge-reconstruction network for the graph-decoder, which allows us to reconstruct multidimensional edge information. This gives the graph-autoencoder the capacity to classify entire graphs, unlike previously existing graph-autoencoders. We use NNConv to incorporate the multidimensional edge and node features as inputs to a graph-autoencoder while utilising edge convolutions to learn inductive latent space representations of QCD jets' graph-structured data. 

The anomaly finder based on the reconstruction loss shows good performance for the non-QCD scenarios that we consider. We further explore the possibility of exploiting latent space variables as discriminants for anomalous jets and find that latent variables can indeed lead to improved anomaly detection by accessing the compressed information of the QCD data. While GNNs are known to be good candidates for trigger-level implementations, we study latent dimension-based anomaly finders with graph-autoencoders. Using latent dimensions instead of the loss has the additional appeal of halving the number of layers, thus resulting in a shallower network. Studying the latent dimension representation of QCD jets therefore provides a compressed arena for new physics discovery by using these observables directly.

\section*{Acknowledgements}
O.A. is supported by the UK Science and Technology Facilities Council (STFC) under grant ST/V506692/1. A.B. and C.E. are supported by the STFC under grant ST/T000945/1. C.E is also supported by the IPPP Associateship Scheme. 
M.S. is supported by the STFC under grant ST/P001246/1. The neural network implementations of this work was performed using the HPC resources (Vikram-100 HPC) and TDP project at PRL.

\appendix
\section{Comparison with Particle Graph Autoencoder}
\label{app:pgae}
\begin{figure*}[th!]
	\centering
	\subfigure[\label{fig:loss_dist_pgae}]{\includegraphics[scale=0.25]{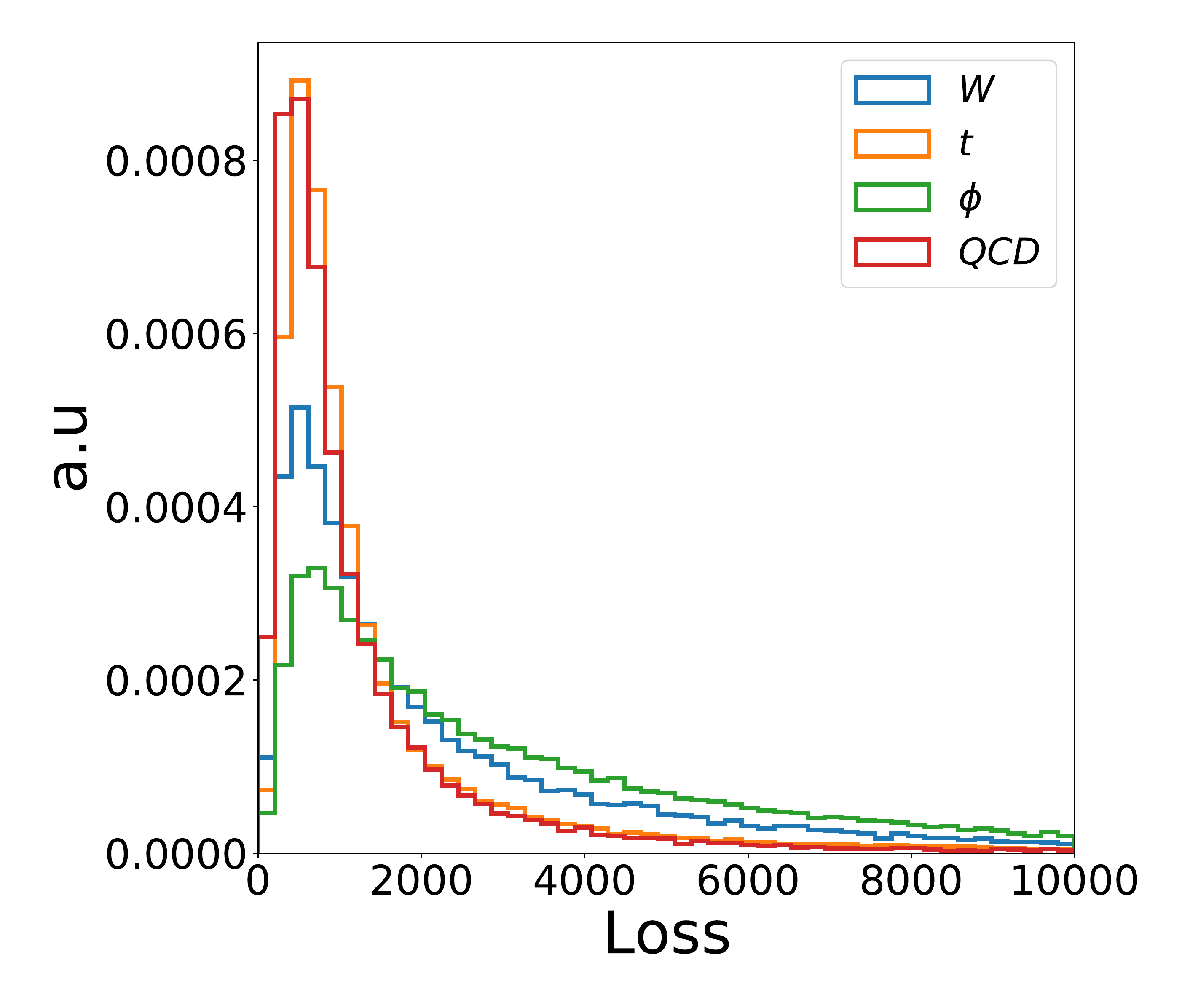}}
	\hskip 0.4cm
	\subfigure[\label{fig:loss_roc_pgae}]{\includegraphics[scale=0.25]{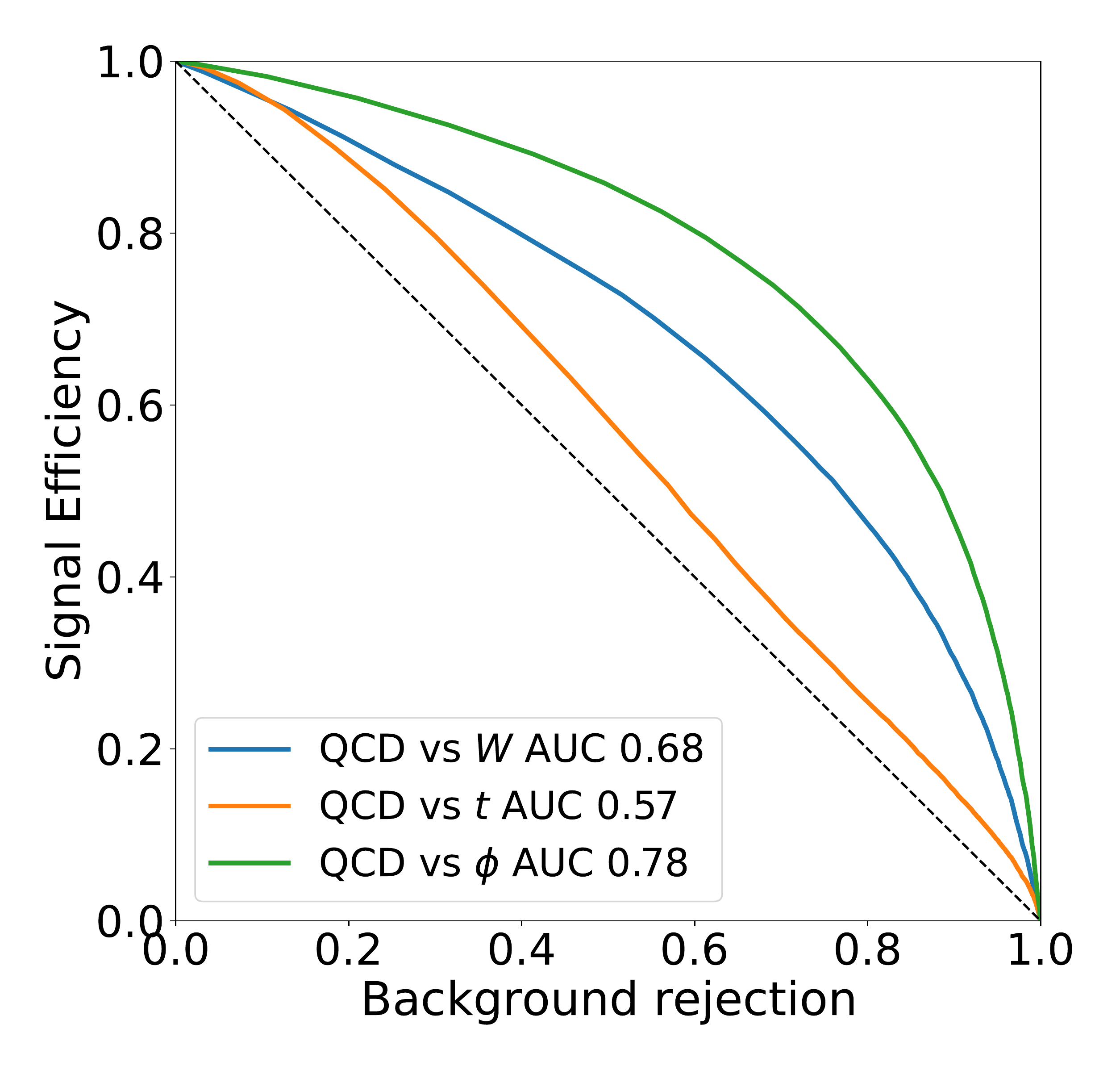}}	
	\caption{Distribution of the loss function of the PGAE (a) and the corresponding ROC curves (b) for the different signal classes for a network trained only on QCD jets.}
	\label{fig:loss_dist_auc_pgae}
\end{figure*}
We compare the performance of our network with the particle graph autoencoders (PGAE) proposed in Ref.~\cite{kasieczka2021lhc}\footnote{We use the code available in \url{https://github.com/stsan9/AnomalyDetection4Jets/tree/rnd_cuts}} with our dataset. This study focussed on identifying anomalous events with dijet signatures (large-radius jets and high $p_T$) and used the two leading jets in the event to learn latent event representations. In contrast, our present focus is jet-level classification. For the input, we consider the four-vector of each microjet as the node feature and use a complete graph with all possible connections. The network is a graph-autoencoder that takes the vectors as input with a single edge convolution to map it to a two-dimensional latent node representation and maps it back with another single edge convolution. We use the mean squared error as the loss function and train with a batch size of 32. For more details of the architecture, we refer the reader to Section 3.7 of Ref.~\cite{kasieczka2021lhc}. We show the distribution of the loss function and the corresponding ROC curve in Fig.~\ref{fig:loss_dist_auc_pgae}. The first thing that we notice is that the location of the peaks is identical for all four classes, with the only difference coming in the tail of the distribution. The value of the AUCs is significantly reduced for $W$ and top jets compared to our work, while for $\phi$-jets, the reduction is not that drastic. Out of the three signal classes, $\phi$ jets are the least QCD-like, and hence, the networks find it easier to distinguish them with less information. At the same time, the edge-reconstruction employed in our architecture helps identify the $W$ and top jets more efficiently.  Hence, we infer that the edge-reconstruction and the multidimensional edge feature representation is crucial for a graph-autoencoder as these complex and physically relevant features are not learned by the network even though they are, in principle, constructed from the node features.  Moreover, using only the node features, the graph autoencoder is insensitive to the n-prong structure of the signals as the AUCs do not follow the usual QCD intuition. The addition of the edge features and their reconstruction enables the graph-autoencoder to learn the signal jets' $n$-prong topology in an unsupervised manner.
\section{Correlation of the loss function with jet variables}
\label{app:loss_corr}

\begin{figure*}[t!]
	\centering
	\includegraphics[scale=0.135]{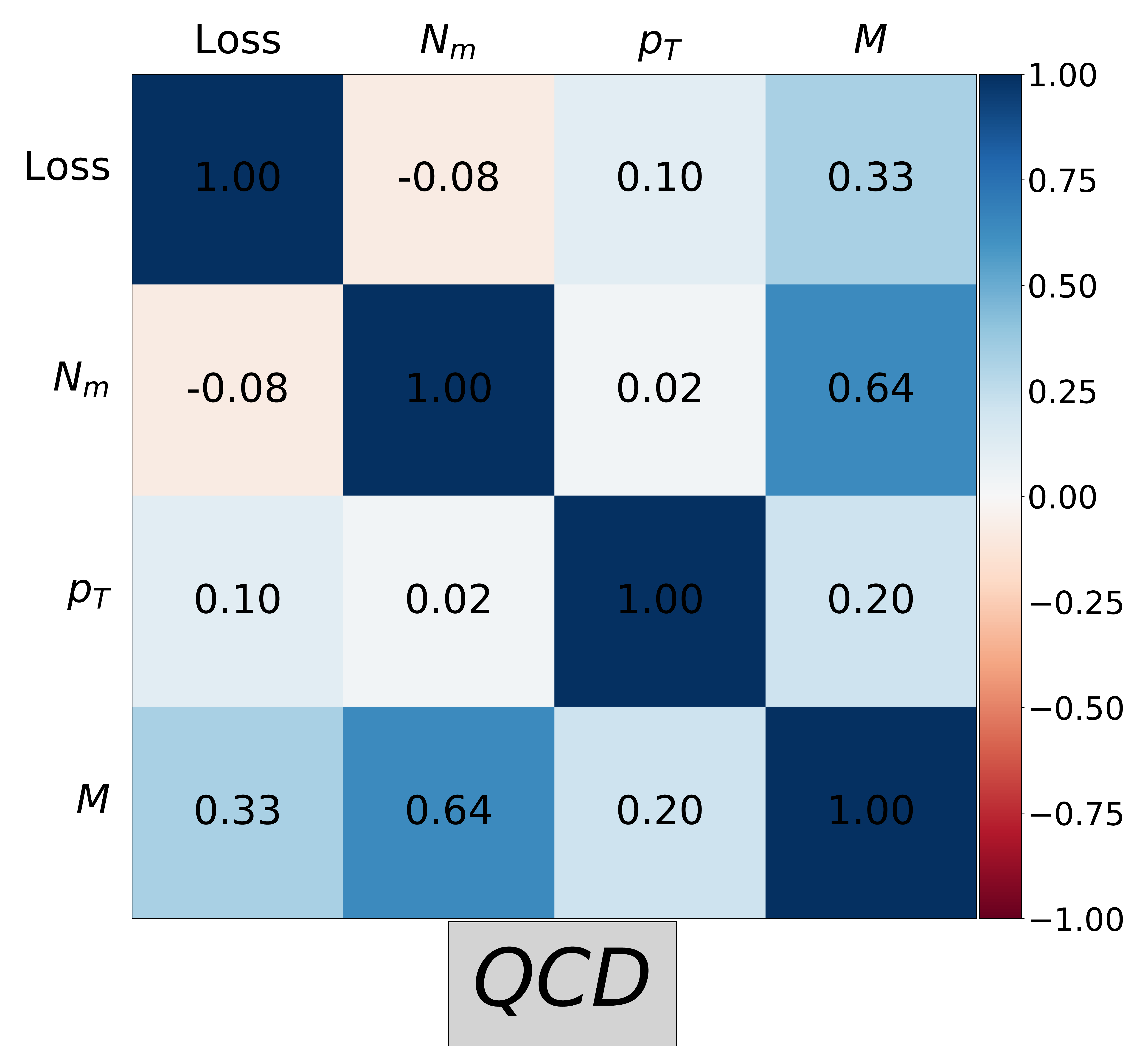}
	\includegraphics[scale=0.135]{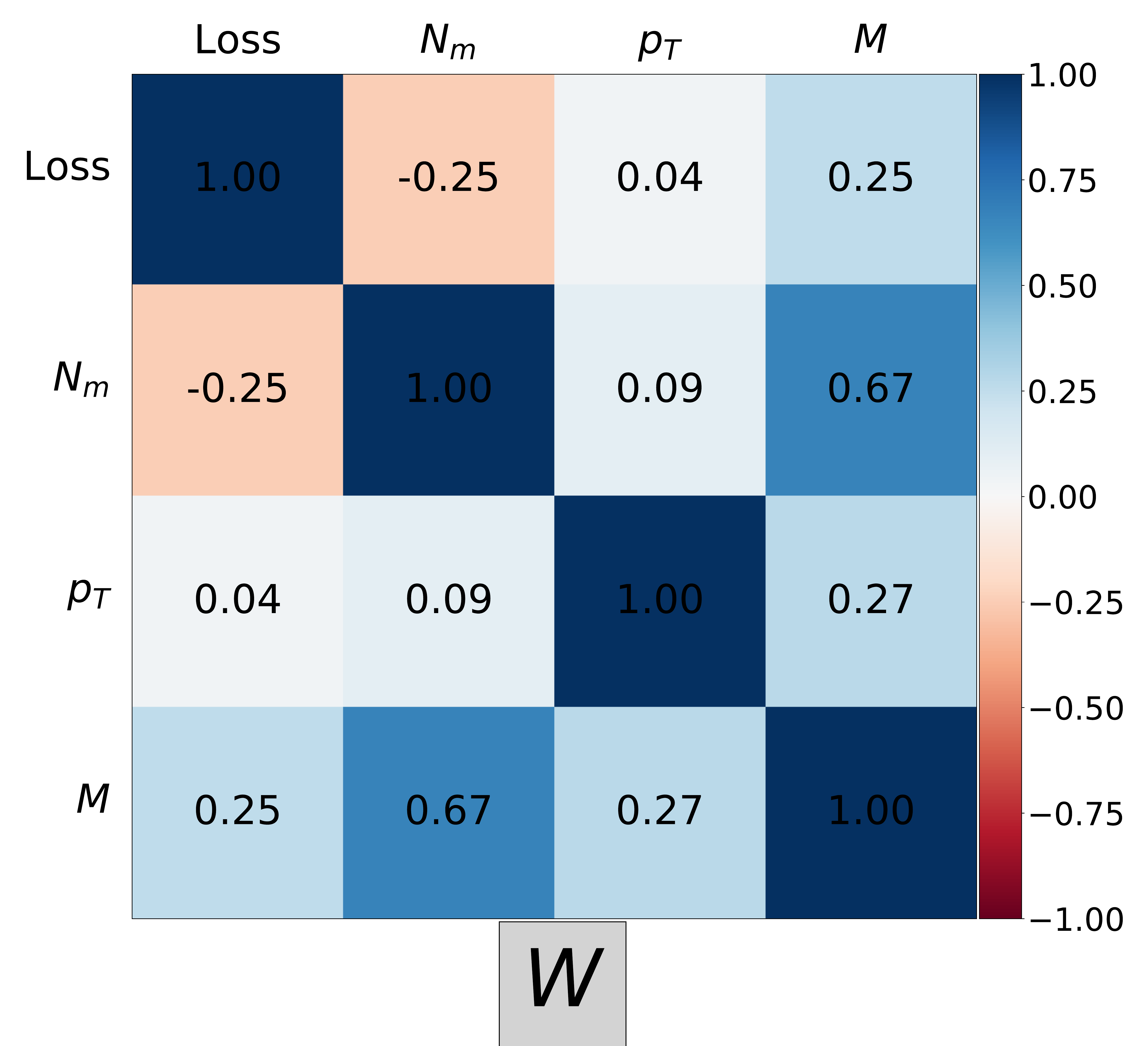}\\
	\includegraphics[scale=0.135]{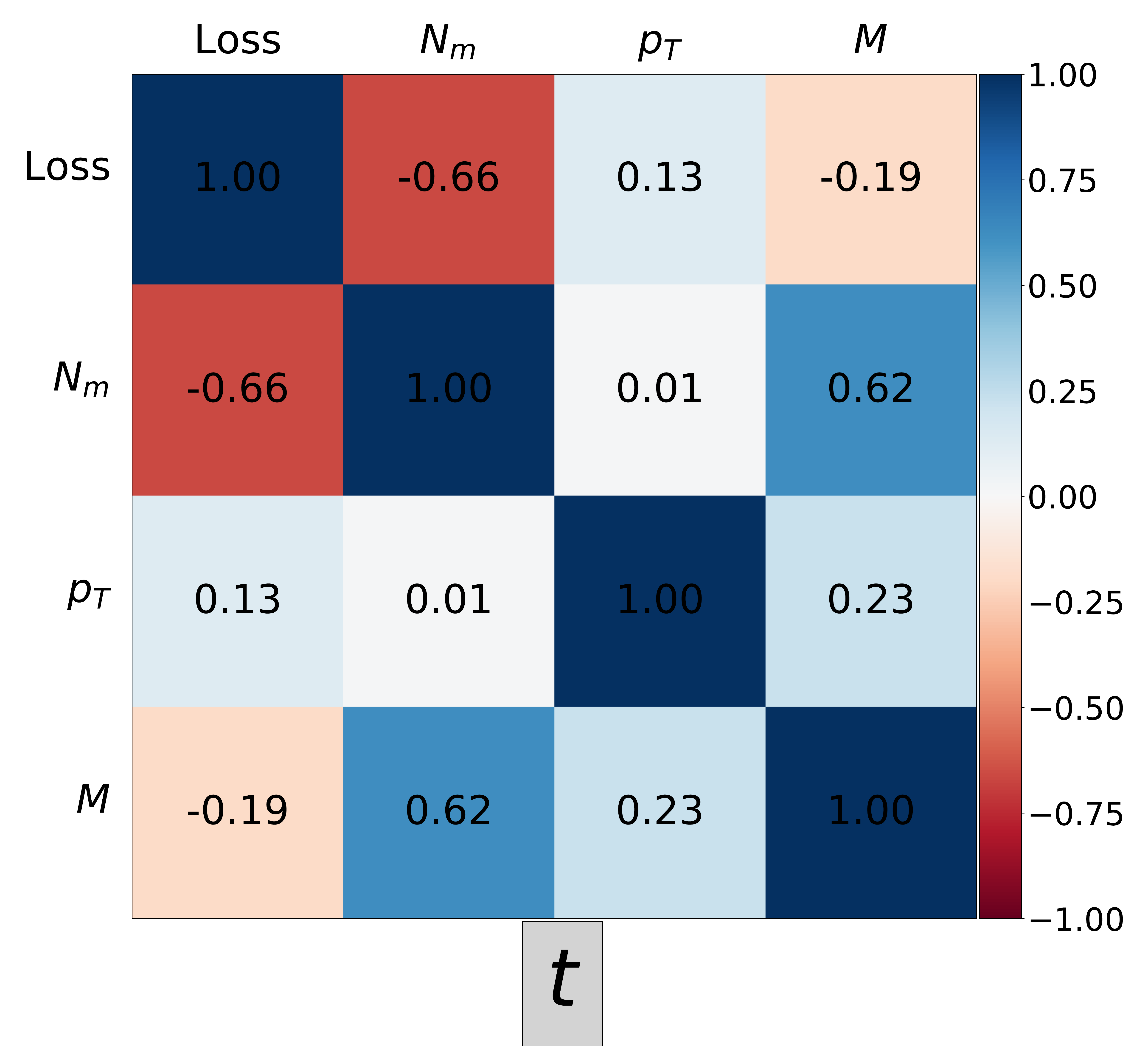}
	\includegraphics[scale=0.135]{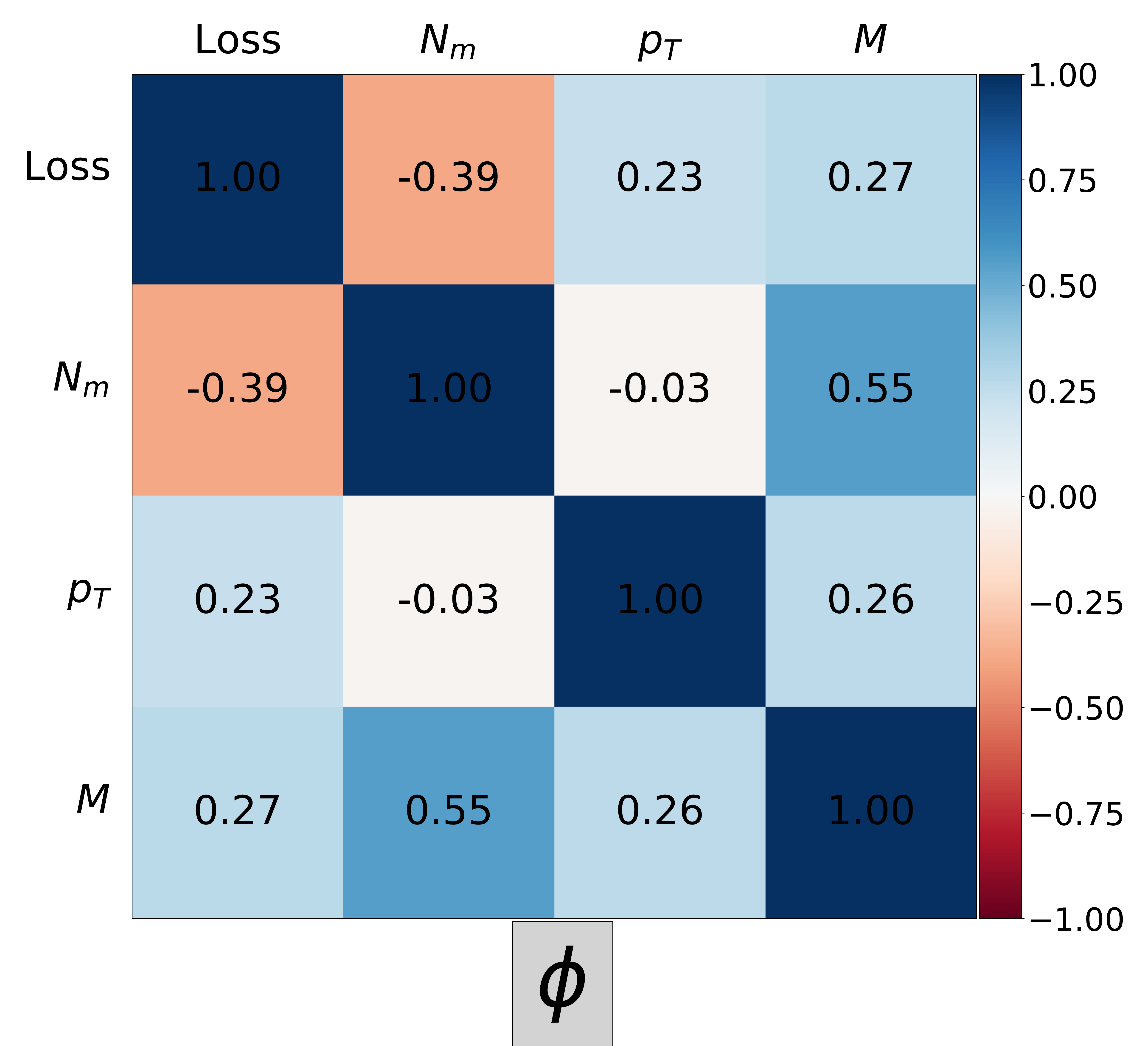}
	\caption{Linear correlation coeeficient between the loss, number of microjets($N_m$), transverse momentum ($p_T$), mass($M$) of the jet for the four jet classes: QCD(top left), W-boson (top right), top-quark (bottom left) and $\phi$(bottom right).}
	\label{fig:loss_corr}
\end{figure*}

The correlation of the loss function with different jet variables is essential in determining the trained network's biases. Although perfectly decorrelated discriminants to the jet's physical variables like transverse momentum ($p_T$), mass ($M$),  or the number of constituents are highly coveted, it is not possible in practice -- known methods to decorrelate them, like adversarial training, diminish the power of the discriminant.  
We discuss the correlation of our network's loss function with the quantities mentioned earlier in this section. The class-wise correlation of the four quantities is shown in Fig.~\ref{fig:loss_corr}. 
We see that the loss function and the $p_T$ are uncorrelated with small positive values (the highest being 0.27 for $\phi$), indicating that the loss function tends to increase with an increase in transverse momentum of the jet slightly, although the increase is minimal for the background QCD jets ($\sim0.10$). Jet mass is an important variable that helps in discriminating different classes of jets. However, a discriminant (the loss function) needs to be decorrelated entirely with jet mass as putting a cut on a correlated variable will lead to artificial bumps in the jet-mass distribution of the selected events. As can be seen, from Fig.~\ref{fig:loss_corr}, the loss function is reasonably correlated with the jet mass even for the QCD jets. Decorrelating the jet mass from the loss can be done via an adversarial network~\cite{Heimel:2018mkt}. 

 The reconstruction efficiency of convolutional autoencoders has been shown to decrease with an increase in the number of non-zero pixels~\cite{Finke:2021sdf}, which leads to the possibility of missing out on potential signals with lower complexities than QCD jets. We find that our network behaves in the opposite way: the reconstruction error reduces with an increase in the number of microjets. More importantly, this reduction is minimal for the QCD jets, suggesting that the network learns a uniform feature of the jet graph regardless of the number of microjets. 
 \begin{figure*}[t!]
 	\centering
 	\includegraphics[scale=0.4]{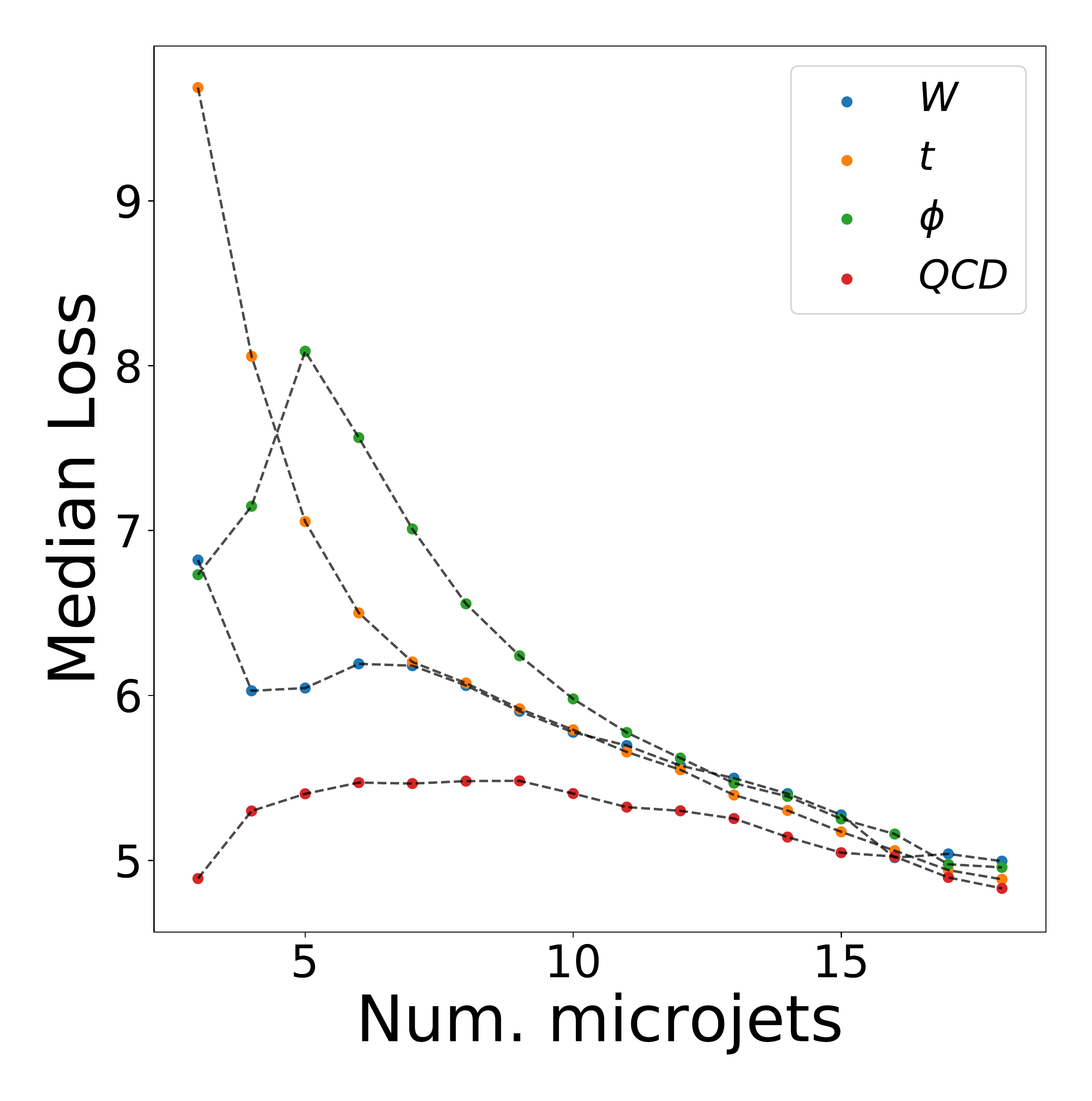}
 	\caption{The median loss of events (from the test dataset) with fixed number of microjets for the various types of jets.}
 	\label{fig:los_mult_med}
 \end{figure*} We can understand this independence via the structure of a graph neural network. A graph convolution layer essentially learns a set of weights shared for all the nodes and edges and hence learns the underlying feature regardless of the number of nodes/edges in the graphs. However, there is a strong negative correlation of the loss of the different signal classes with the number of microjets which can be understood via the fact that any extra radiation other than the said multiplicities essentially arise from QCD splittings. To further understand this behavior, we plot the median loss of the events with a fixed number of microjets for the four classes in Fig.
\ref{fig:los_mult_med}. The initial increase in the median loss from three to five for the four-pronged $\phi$ jets further solidifies the preceding argument regarding the decrease of the loss function with an increase in the microjet multiplicity. Such a peak is absent for the lower multiplicity signal classes. 
      
\bibliographystyle{JHEP}
\bibliography{references.bib}

\end{document}